\documentclass[numberedappendix,revtex4]{emulateapj}

\newcommand{\myemail}{\url{bld002@email.uark.edu}}
\newcommand {\apgt} {\ {\raise-.5ex\hbox{$\buildrel>\over\sim$}}\ }
\newcommand {\aplt} {\ {\raise-.5ex\hbox{$\buildrel<\over\sim$}}\ }

\shorttitle{BHMF Derived from Local Spiral Galaxies}
\shortauthors{Davis et al.}

\usepackage{color}
\usepackage{soul}
\definecolor{gray}{gray}{0.5}
\usepackage{url}
\usepackage{apjfonts}
\usepackage[latin1]{inputenc}
\usepackage{tikz}
\usetikzlibrary{shapes,arrows}
\usepackage{epsfig}
\usepackage{amsmath}
\usepackage{wasysym}

\submitted{Published in The Astrophysical Journal, 789:124 (16pp), 2014 July 10}

\begin{document}

\title{The Black Hole Mass Function Derived from Local Spiral Galaxies}
      
\author{Benjamin L. Davis\altaffilmark{1}, Joel C. Berrier\altaffilmark{1,2,5}, Lucas Johns\altaffilmark{3,6}, Douglas W. Shields\altaffilmark{1,2}, Matthew T. Hartley\altaffilmark{2}, Daniel Kennefick\altaffilmark{1,2}, Julia Kennefick\altaffilmark{1,2}, Marc S. Seigar\altaffilmark{1,4}, and Claud H. S. Lacy\altaffilmark{1,2}}
\altaffiltext{1}{Arkansas Center for Space and Planetary Sciences, University of Arkansas, 346 1/2 North Arkansas Avenue, Fayetteville, AR 72701, USA; \myemail}
\altaffiltext{2}{Department of Physics, University of Arkansas, 226 Physics Building, 835 West Dickson Street, Fayetteville, AR 72701, USA}
\altaffiltext{3}{Department of Physics, Reed College, 3203 SE Woodstock Boulevard, Portland, OR, 97202, USA}
\altaffiltext{4}{Department of Physics and Astronomy, University of Arkansas at Little Rock, 2801 South University Avenue, Little Rock, AR 72204, USA}
\altaffiltext{5}{Now at Department of Physics and Astronomy, Rutgers, The State University of New Jersey, 136 Frelinghuysen Road, Piscataway, NJ 08854-8019, USA}
\altaffiltext{6}{Now at Department of Physics, University of California, San Diego, La Jolla, CA 92092, USA}

\begin{abstract}

We present our determination of the nuclear supermassive black hole mass (SMBH) 
function for spiral galaxies in the local universe, established from a 
volume-limited sample consisting of a statistically complete collection 
of the brightest spiral galaxies in the southern ($\delta  < 0^{\circ}$) 
hemisphere. 
Our SMBH mass function agrees well at the high-mass end with previous values
given in the literature. At the low-mass end, inconsistencies exist 
in previous works that still need to be resolved, but our work is more
in line with expectations based on modeling of black hole evolution. 
This low-mass end of the spectrum is critical to our understanding
of the mass function and evolution of black holes since the epoch of
maximum quasar activity.
A limiting luminosity 
(redshift-independent) distance, $D_{\rm L} = 25.4$ Mpc ($z = 0.00572$) 
and a limiting absolute $B$-band magnitude, $\mathfrak{M}_{\rm B} = -19.12$ define the sample. 
These limits define a sample of 140 spiral galaxies, with 128 measurable 
pitch angles to establish the pitch angle distribution for this sample. 
This pitch angle distribution function may be useful in the study of
the morphology of late-type galaxies.
We then use an established relationship between the logarithmic spiral arm 
pitch angle and the mass of the central SMBH in a host 
galaxy in order to estimate the mass of the 128 respective SMBHs in this volume-limited sample. This result effectively gives 
us the distribution of mass for SMBHs residing in spiral 
galaxies over a lookback time, $t_{\rm L} \leq 82.1$ $h_{67.77}^{-1}$ 
Myr and contained within a comoving volume, $V_{\rm C} = 3.37$ $\times$ 
$10^4$ $h_{67.77}^{-3}$ Mpc$^3$. We estimate that the density of SMBHs residing in spiral galaxies in the local universe is $\rho = 
5.54_{-2.73}^{+6.55}$ $\times$ $10^4$ $h_{67.77}^3$ $M_{\odot}$ Mpc$^{-3}$. Thus, our derived cosmological SMBH mass density for spiral galaxies is $\Omega_{\rm BH} = 4.35_{-2.15}^{+5.14}$ $\times$ $10^{-7}$ $h_{67.77}$. Assuming that black holes grow via baryonic accretion, we predict that $0.020_{-0.010}^{+0.023}$ $h_{67.77}^3$ $\permil$ of the universal baryonic inventory ($\Omega_{\rm BH}/\omega_{\rm b}$) is confined within nuclear SMBHs at the center of spiral galaxies.
\end{abstract}

\keywords{black hole physics --- cosmology: miscellaneous --- galaxies: evolution --- galaxies: spiral}

\section{Introduction}

Strong evidence suggests that supermassive black holes (SMBHs) reside in the
nuclei of most galaxies and that correlations exist between the
mass of the SMBH and certain properties of the host galaxy
\citep{Kormendy:Richstone:1995,Kormendy:Gebhardt:2001}. It is therefore
possible to conduct a census by studying the numerous observable galaxies
in our universe in order to estimate demographic information (i.e., mass)
for the population of SMBHs in our universe. Following the discovery of
quasars \citep{Schmidt:1963} and the early suspicion that their power
sources were in fact SMBHs \citep{Salpeter:1964,Lynden-Bell:1969}, the study
of quasar evolution via quasar luminosity functions (QLFs) has resulted in
notable successes in understanding the population of SMBHs in the universe
and their mass function.
But, studies of the supermassive black hole mass function (BHMF)
have left us with no clear consensus, especially
at the low-mass end of the spectrum.

It is of particular interest to understand the low-mass end of the BHMF
in order to understand how
the QLF of past epochs evolves into the BHMF 
of today \citep{Shankar:2009b}.
It is now widely accepted that black holes in active galactic nuclei (AGN) do not generally
accrete at the Eddington limit \citep{Shankar:2009b}.
This is not a problem for the brightest and most visible AGN, presumably powered by large black holes accreting
at a considerable fraction of their Eddington limit.
Smaller black holes cannot imitate this luminosity without accreting
at super-Eddington rates.
However, one cannot
know whether a relatively dim quasar contains a small black hole
accreting strongly or a larger black hole accreting at a relatively low rate
(a small fraction of its Eddington limit). It is therefore not easy to tell
what the BHMF was for AGN in the past, because it is non-trivial to
count the number of lower-mass black holes (those with masses in the range
of less than a million to ten million solar masses). 
However, if we counted the number
of local lower-mass black holes, the requirement that the BHMF from
the quasar epochs evolve into the local BHMF could significantly constrain
the BHMF in the past, as well as determine a more complete local picture.
Thus, one should pay attention to late-type (spiral) galaxies,
since a significant fraction of these lower-mass black holes are found
in such galaxies (our own Milky Way being an example).

Some indicators of SMBH mass such as the central stellar velocity 
dispersion \citep{Gebhardt:2000a,Ferrarese:Merritt:2000} or S\'ersic index 
\citep{Graham:Driver:2007} have been used to construct
BHMFs for
early-type galaxies \citep[e.g.,][]{Graham:2007}. They have been
used also to study late-type galaxies, but not always with success
because these quantities 
are defined for the bulge component of galaxies, measuring them 
in disk galaxies 
requires decomposition into separate components of the galactic bulge, disk, 
and bar.
Thus, we are currently handicapped in the
study of the low-mass end of the BHMF by the relative scarcity of
information on the mass function of spiral galaxies.
One approach has been to use luminosity or other functions available
for all galaxy types in a sample to produce a mass function based upon
the relevant scaling relation \citep{Salucci:1999,Aller:Richstone:2002,Shankar:2004,Shankar:2009,Tundo:2007}.
Our approach contrasts with this one by taking individual measurements
of a quantity for each galaxy individually in a carefully selected
and complete local sample.

Recently, it has been shown that there is a strong correlation between
SMBH mass and spiral arm pitch angle in disk galaxies
\citep{Seigar:2008,Berrier:2013}. This correlation presents a number
of potential advantages for the purposes of developing the BHMF
at lower masses. First, there is evidence that it has lower scatter when applied to disk galaxies than
any of the other correlations that have been presented \citep{Berrier:2013}. Second, the pitch angle is less problematically
measured in disk galaxies than the other features, which is likely the explanation for the lower scatter. It does not require 
any decomposition of the bulge, disk, or bar components besides a trivial
exclusion of the central region of the galaxy before the analysis (described
below in \S\ref{Pitch_Method}). Finally, it can be derived from imaging data alone, which is already
available in high quality for many nearby galaxies. 

It may be objected that the spiral arm structure of a disk galaxy spans
tens of thousands of light years, many orders of magnitude greater than
the scale (some few light years) over which the SMBH
is the dominating influence at the center of a galaxy. However, as with other
correlations of this type, the spiral arm pitch angle does not directly
measure the black hole mass, rather it is a measure of the mass of the
central region of the galaxy (the bulge in disk-dominated galaxies). 
The modal density wave theory \citep{Lin:Shu:1964} describes the spiral arm structure
as a standing wave pattern created by density waves propagating through the 
disk of the galaxy. The density waves are generated by resonances between 
orbits at certain radii in the disk. As with other standing wave patterns, 
the wavelength, and therefore the pitch angle of the spiral arms, depends
on a ratio of the mass density in the disk to the ``tension'' provided by
the central gravitational well, and thus to the mass of the galaxy's
central region. In the case of spiral density waves in Saturn's rings, the
dependence of the pitch angle on the ratio of the disk mass density to the
mass of the central planet has been conclusively shown \citep{Shu:1984}. In
galaxies, the central bulge provides (in most cases) the largest part of
this central mass. Since it is well known that the mass of the central SMBH
correlates with the mass of the central bulge component, it is not at all
surprising to find that it also correlates with the spiral arm pitch angle
(further details can be found in \citealt{Berrier:2013}).

The pitch angle ($P$) of the spiral arms 
of a galaxy is inversely proportional to the mass of the central bulge of a galaxy; specifically 
\begin{equation}
\cot \left | P \right | \propto M_{\rm Bulge},
\end{equation}
where $M_{\rm Bulge}$ is the bulge mass of the galaxy. This is a requirement of all current 
theories regarding the origin of a spiral structure in galaxies. Since the bulge mass is directly 
proportional to the velocity dispersion of the bulge via the virial theorem, i.e.,
\begin{equation}
\sigma^{2} \approx \frac{GM_{\rm Bulge}}{R},
\end{equation}
where $G$ is the universal gravitational constant and $R$ is the radius of the bulge; and the 
nuclear SMBH mass is directly proportional to the velocity dispersion via the $M$--$\sigma$ relation, i.e.,
\begin{equation}
M \propto \sigma^{\alpha},
\end{equation}
with $\alpha = 4.8 \pm 0.5$ \citep{Ferrarese:Merritt:2000}; it therefore follows that the mass ($M$)
of the nuclear SMBH must be indirectly proportional to the pitch angle of its host galaxy's spiral arms, i.e.,
\begin{equation}
M \propto 10^{-(0.062 \pm 0.009) \left | P \right |},
\end{equation}
as shown in Equation (\ref{M-P_Relation}).

Admittedly, galaxies are complex structures. However, a number of measurable features of disk galaxies
are now known to correlate with each other, even though they are measured on
very different length scales (e.g., $\sigma$, bulge luminosity, S\'ersic Index, and
spiral arm pitch angle). Each of these quantities is influenced, or even determined,
by the mass of the central
bulge of the disk galaxy, and this quantity in turn seems to correlate quite well with the
mass of the central black hole. The precise details of how this nexus of what
we might call ``traits'' of the host galaxy correlate to the black hole
mass is still subject to debate (see for instance \citet{Lasker:2014},
which shows that central black hole may correlate equally well with total
galaxy luminosity as with central bulge luminosity). Nevertheless, what
seems to link the various galaxy ``traits'' (such as pitch angle, $\sigma$,
and so on) is that they are all measures of the mass in the central regions
of the galaxy.

That this hidden feature of galaxies, the black hole mass,
should be indirectly estimable from measurements of highly visible morphological features,
such as pitch angle, is a considerable boon to astronomers. Pitch angle, as a marker for
black hole mass, has a number of distinct advantages over other possible markers. It is
obtainable from imaging data alone. It is quite unambiguous for many spiral galaxies,
whereas other quantities, such as $\sigma$ or S\'ersic index, depend upon the astronomer's
ability to disentangle bulge components from bar and disk components. Finally, while $\sigma$
or stellar velocity dispersion depends on the size of the slit used in spectroscopy, with one particular
size giving the desired correlation with black hole mass, pitch angle can be considered
relatively constant for any annulus-shaped portion of the disk (as long as the spiral arm
pattern is truly logarithmic, which is usually the case for all but the very outermost part
of the disk). This combination of advantages may permit pitch angle to be used on even
larger samples in the future, yielding a better understanding of the evolution of the black hole mass
function and its properties in different parts of the universe.


It is worth mentioning the point made by \citet{Kormendy:2011,Kormendy:Ho:2013} that
the $M$--$\sigma$ relation may not work at all for spiral galaxies with
pseudo-bulges rather than classical bulges. This viewpoint has
been controversial \citep[e.g.,][]{Graham:2011a}, but it is born out of
the observation that $\sigma$ is defined with ``hot'' bulges rather
than pseudo-bulges in mind in the first place.
It can be observed that density wave theory still expects that pitch
angle should depend on the central mass of the galaxy, regardless of whether or
not the galaxy has a bulge or pseudo-bulge \citep{Roberts:1975}. Unfortunately, it is not always trivial to determine which spirals
have pseudo-bulges, but it is worth noting that four of the sample used in
defining the $M$--$P$ relation in \citet{Berrier:2013} are specifically classified by \citet{Kormendy:2011} as pseudo-bulges. In addition, \citet{Kormendy:2011} feel that a S\'ersic index
of two can be a good indication that a galaxy has a pseudo-bulge. \citet{Berrier:2013}
report S\'ersic indices for the majority of the galaxies used in their determination
of the $M$--$P$ relation and roughly half of them have S\'ersic indices less than two.
Thus, there are some grounds for expecting that
the $M$--$P$ relation may work about as well for pseudo-bulges as for
galaxies with classical bulges.


In this paper, we present our determination of the BHMF for local spiral 
galaxies. We conducted our analysis from a statistically complete sample 
of local spiral galaxies by measuring their pitch angles using the method 
of \citet{Me:2012} and use the well-established $M$--$P$ relation 
\citep{Berrier:2013} to convert the pitch angles ($P$) to 
SMBH masses ($M$). 
The paper is outlined as follows. \S\ref{Methodology} discusses the importance of spiral galaxies, our methodology for measuring pitch angles, and presents the $M$--$P$ relation as found by \citet{Berrier:2013}. \S\ref{Sect_Sample} 
details our volume-limited sample of spiral galaxies. \S\ref{Sect_Pitch_Dist} discusses 
the results of our pitch angle measurements and their resulting distribution. 
\S\ref{Sect_Mass_Dist} details the conversion of our pitch angle distribution 
to a black hole mass distribution. \S\ref{Sect_BHMF} reveals our BHMF for 
spiral galaxies. 
\S\ref{Sect_Discussion} provides a discussion on the implication of 
our results. 
Finally, \S\ref{Conclusions} contains concluding remarks and a summary of results. We also include in the Appendix, sections discussing the pitch angle of the Golden Spiral (\S\ref{Golden Appendix}), the Milky Way (\S\ref{MW Appendix}), and listing details of our full sample (\S\ref{Appendix Sample}). Throughout this paper, we adopt a $\Lambda$CDM (Lambda-Cold Dark Matter) 
cosmology with the best-fit {\it Planck}+WP+highL+BAO cosmographic parameters 
estimated by the {\it Planck} mission \citep{Planck:2013}: $\omega_{\rm b} = 0.022161$, $\Omega_{\rm M} = 0.3071$, 
$\Omega_\Lambda = 0.6914$, and $h_{67.77}  = h/0.6777 = H_0/(67.77$ km s$^{-1}$ Mpc$^{-1}) \equiv 1$.

\section{Methodology}\label{Methodology}

Our goal of assembling a BHMF for the local universe is accomplished by using pitch angle measurements to estimate black hole masses. Using a well-defined sample, we can construct a representative BHMF. We have completed pitch angle measurements for a volume-limited set of local spiral galaxies, with the aim of ultimately determining the BHMF for the local universe,
\begin{equation}
\frac{\partial{N}}{\partial{M}} = \frac{\partial{N}}{\partial{P}}\frac{\partial{P}}{\partial{M}},
\label{eqn4}
\end{equation} 
where $N$ is the number of galaxies and $M$ is SMBH mass. The pitch angle measurements for the volume-limited sample give us $\frac{\partial{N}}{\partial{P}}$, while $\frac{\partial{P}}{\partial{M}}$ for spiral galaxies in the local universe has already been discussed and evaluated in the literature \citep{Seigar:2008,Berrier:2013}.

\subsection{How We Measure Pitch Angle}\label{Pitch_Method}

The best geometric measure for logarithmic spirals is the pitch angle, and this can be measured for any galaxy in which a spiral structure can be discerned, independently of the distance to the galaxy \citep{Me:2012}. We measure galactic logarithmic spiral arm pitch angle by implementing a modified two-dimensional (2D) fast Fourier transform (FFT) software called {\it 2DFFT} to decompose charge-coupled device (CCD) images of spiral galaxies into superpositions of logarithmic spirals of different pitch angles and numbers of arms, or harmonic modes ($m$). Galaxies with random inclinations between the plane of their disk and the plane of the sky are deprojected to a face-on orientation. Although \citet{Ryden:2004} has argued that disk galaxies are inherently non-circular in outline, their typical ellipticity is not large and, as has been shown by \citet{Me:2012}, a small ($\lessapprox10^{\circ}$ error in inclination angle) departure from circularity does not adversely affect the measurement of the pitch angle. From a user-defined measurement annulus centered on the center of the galaxy, pitch angles are computed for all combinations of measurement annuli, where the inner radius is made to vary by consecutive increasing integer pixel values from zero to one less than the selected outer radius. The pitch angle corresponding to the frequency with the maximum amplitude is captured for the first six non-zero harmonic modes (i.e., for spiral arm patterns containing up to six arms). A mean pitch angle for a galaxy is found by examining the pitch angles measured for different inner radii, selecting a sizable radial region over which the pitch angle is stable. The error depends mostly on the amount of variation in the pitch angle over this selected region. Full details of our methodology for measuring galactic logarithmic spiral arm pitch angle via 2D FFT decomposition can be found in \citet{Me:2012}. 

\subsection{The $M$--$P$ Relation}\label{Joel}

The pitch angle of a spiral galaxy has been shown to correlate well with the mass of the central SMBH residing in that galaxy \citep{Berrier:2013}. Thus, using the linear best-fit $M$--$P$ relation established by \citet{Berrier:2013} for local spiral galaxies,
\begin{equation}
\log(M/M_{\odot}) = (b \pm \delta b) - (k \pm \delta k)\left | P \right |,
\label{M-P_Relation}
\end{equation}
with $b = 8.21$, $\delta b = 0.16$, $k = 0.062$, and $\delta k = 0.009$, we can estimate the SMBH masses for a sample of local spiral galaxies merely by measuring their pitch angles using the method of \citet{Me:2012}. The linear fit of \citet{Berrier:2013} has a reduced $\chi^2 = 4.68$ with a scatter of $0.38$ dex, which is lower than the intrinsic scatter ($\Delta = 0.53 \pm 0.10$ dex) of the $M$--$\sigma$ relation for late-type galaxies \citep{Gultekin:2009} and the rms residual (0.90 dex) for the SMBH mass--spheroid stellar mass relation for S\'ersic galaxies \citep{Scott:2013} in the $\log{M}$ direction. Ultimately, by determining the product of the mass distribution and the pitch angle distribution of a sample with a given volume, we may construct a BHMF for local late-type galaxies.

\section{Data}\label{Sect_Sample}

\begin{figure} 
\includegraphics[trim = 15mm 0mm 17mm 6mm, clip,width=\columnwidth]{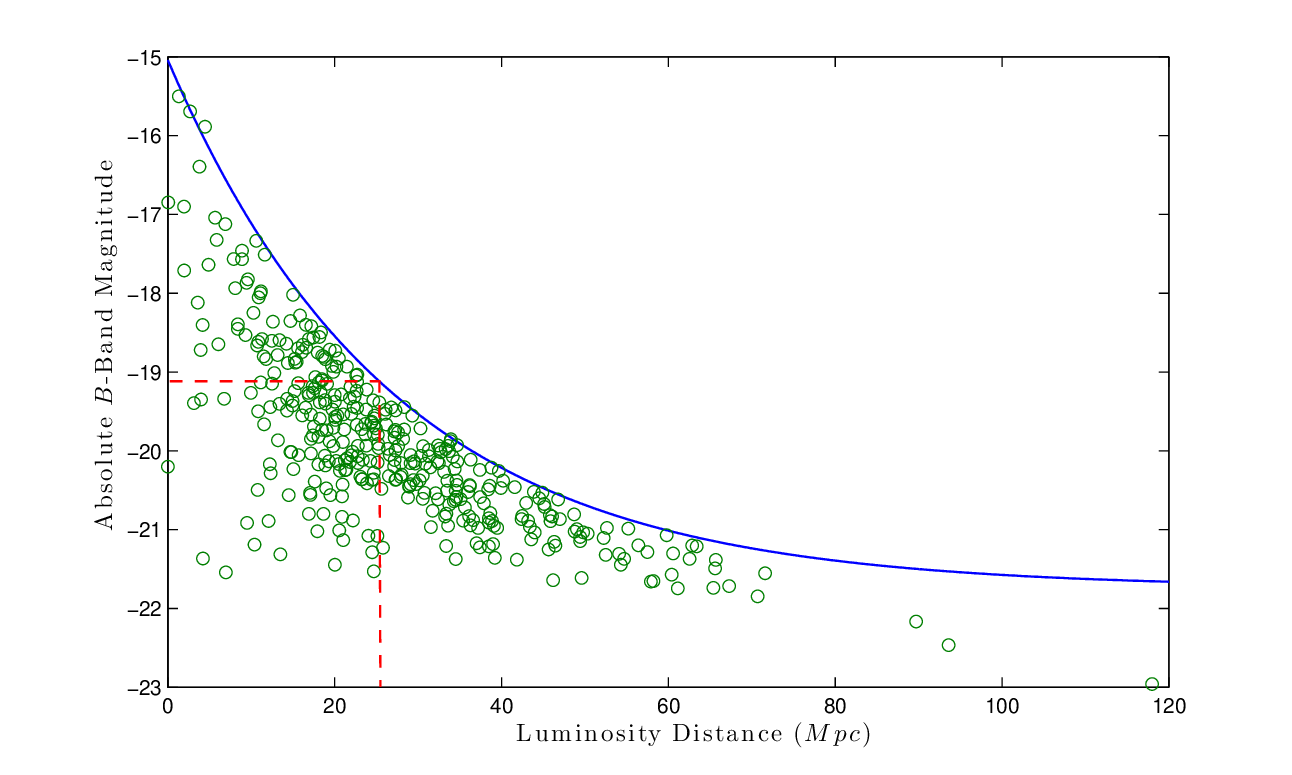}
\includegraphics[trim = 16mm 0mm 17mm 6mm, clip,width=\columnwidth]{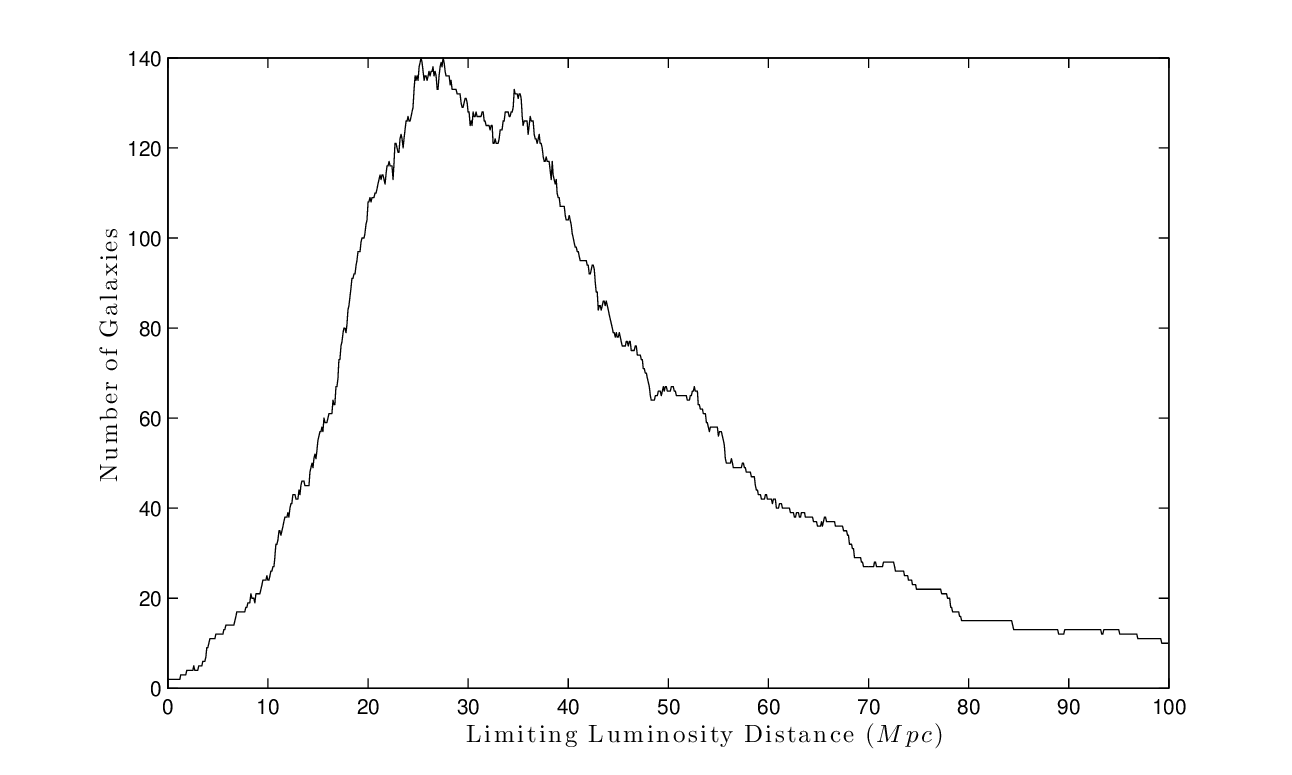}
\caption{Top: luminosity distance vs. absolute $B$-band magnitude for all of the spiral galaxies (385) found using the magnitude-limiting selection criteria ($B_{\rm T} \leq 12.9$ and $\delta < 0^{\circ}$). The upper limit absolute magnitude can be modeled as an exponential and is plotted here as the solid \textcolor{blue}{blue} line. The dashed \textcolor{red}{red} rectangle is constructed to maximize the number of galaxies in the volume-limited sample. The limiting luminosity distance and absolute $B$-band magnitude are set to be $25.4$ Mpc and $-19.12$, respectively. Bottom: histogram showing the number of galaxies contained in the box in the top panel as the box is allowed to slide to new positions based on the limiting luminosity distance. Note there is a double peak in the histogram maximizing the sample each at 140 galaxies. The two possible combinations are $D_{\rm L} = 25.4$ Mpc and $\mathfrak{M}_{\rm B} = -19.12$ or $D_{\rm L} = 27.6$ Mpc and $\mathfrak{M}_{\rm B} = -19.33$. We chose to use the former (leftmost peak) because its volume-limiting sample is complete for galaxies with dimmer intrinsic brightness. In total, the two samples differed by only 20 non-mutual galaxies, a difference of $\approx 14\%$. Complete volume-limited samples were computed for limiting luminosity distances ranging from $0.001$ Mpc to $100.000$ Mpc in increments of $0.001$ Mpc.\label{Vol-limit}}
\end{figure}

In order to quote a meaningful BHMF, it is first necessary to identify an appropriate sample of host galaxies. We have elected to pursue a volume-limited sample; that is, a population of host galaxies that are contained within a defined volume of space and are brighter than a limiting luminosity. For the sake of defining a statistically complete, magnitude-limited sample, we select southern hemisphere ($\delta < 0^{\circ}$) galaxies with a magnitude limit, $B_{\rm T} \leq 12.9$, based on the Carnegie-Irvine Galaxy Survey (CGS); this results in 605 galaxies \citep{Ho:2011}. Our sample is selected from galaxies included in the {\it CGS} sample, because it is a very complete sample of nearby galaxies for which excellent imaging is freely available (we used a small
number of {\it CGS} images, whose pitch angles were previously reported in
\citealt{Me:2012}, other images were obtained from the NASA/IPAC Extragalactic Database (NED)). Using this as our parent sample plus the Milky Way gives us a total of 385 spiral galaxies; we then select only spiral galaxies within a volume-limited sample defined by a limiting luminosity (redshift-independent) distance\footnote{The mean redshift-independent distance averaged from all available sources listed in the {\it NED}, \url{http://ned.ipac.caltech.edu/forms/d.html}}, $D_{\rm L} = 25.4$ Mpc ($z = 0.00572$) and a limiting absolute $B$-band magnitude, $\mathfrak{M}_{\rm B} = -19.12$ (see Figure \ref{Vol-limit}).
This results in a volume-limited sample of 140 spiral galaxies within a region of space with a comoving volume, $V_{\rm C} = 3.37$ $\times$ $10^4$ $h_{67.77}^{-3}$ Mpc$^3$ and a lookback time, $t_{\rm L} \leq 82.1$ $h_{67.77}^{-1}$ Myr. The dimmest (absolute magnitude) and most distant galaxies included in the volume-limited sample are PGC 48179 ($\mathfrak{M}_{\rm B} = -19.12$) and IC 5240 ($D_{\rm L} = 25.4$ Mpc), respectively.

\begin{figure} 
\includegraphics[trim = 4mm 0mm 11mm 6mm, clip,width=\columnwidth]{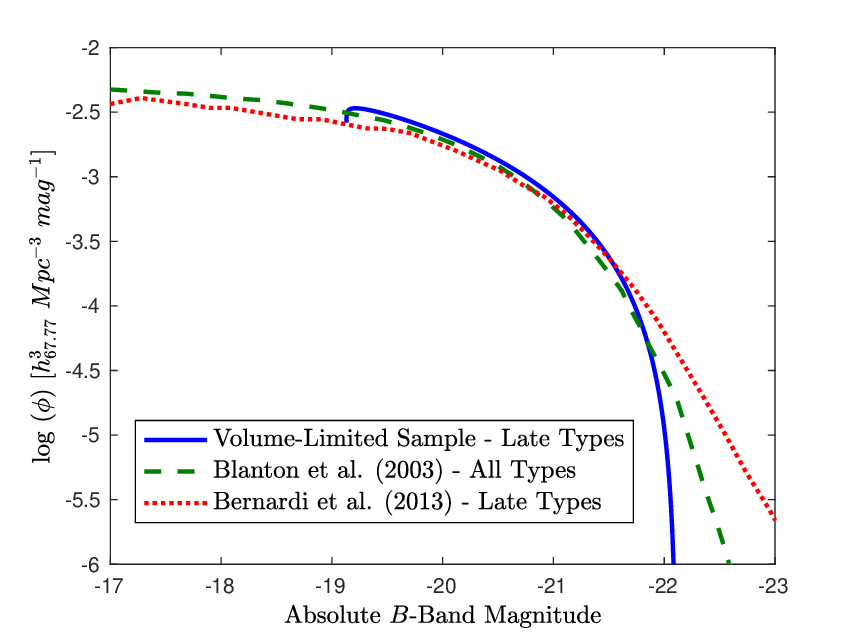}
\caption{Luminosity function for the 140 member, volume-limited sample of galaxies obtained from the larger CGS sample. The function is given here (solid \textcolor{blue}{blue} line) in terms of the probability density function, fit to the results of Equation (\ref{LF_Eqn}). The function abruptly stops on the dim end due to our exclusion of galaxies with $\mathfrak{M}_{\rm B} > -19.12$. Superimposed for comparison is the $r$-band luminosity functions of $z \approx 0.1$ galaxies selected from the Sloan Digital Sky Survey (SDSS) for all galaxy types \citep{Blanton:2003} and late types \citep{Bernardi:2013}; illustrated as \textcolor[rgb]{0,0.5,0}{green} dashed and \textcolor{red}{red} dotted lines, respectively. These have all been shifted by $B - r = 0.67$ mag, the average color of an Sbc spiral \citep{Fukugita:1995}, which is roughly the median Hubble type of both the CGS and our derivative volume-limited sample.\label{Lum_Fcn}}
\end{figure}

\begin{figure}[b] 
\includegraphics[trim = 0mm 0mm 0mm 0mm, clip, width=\columnwidth]{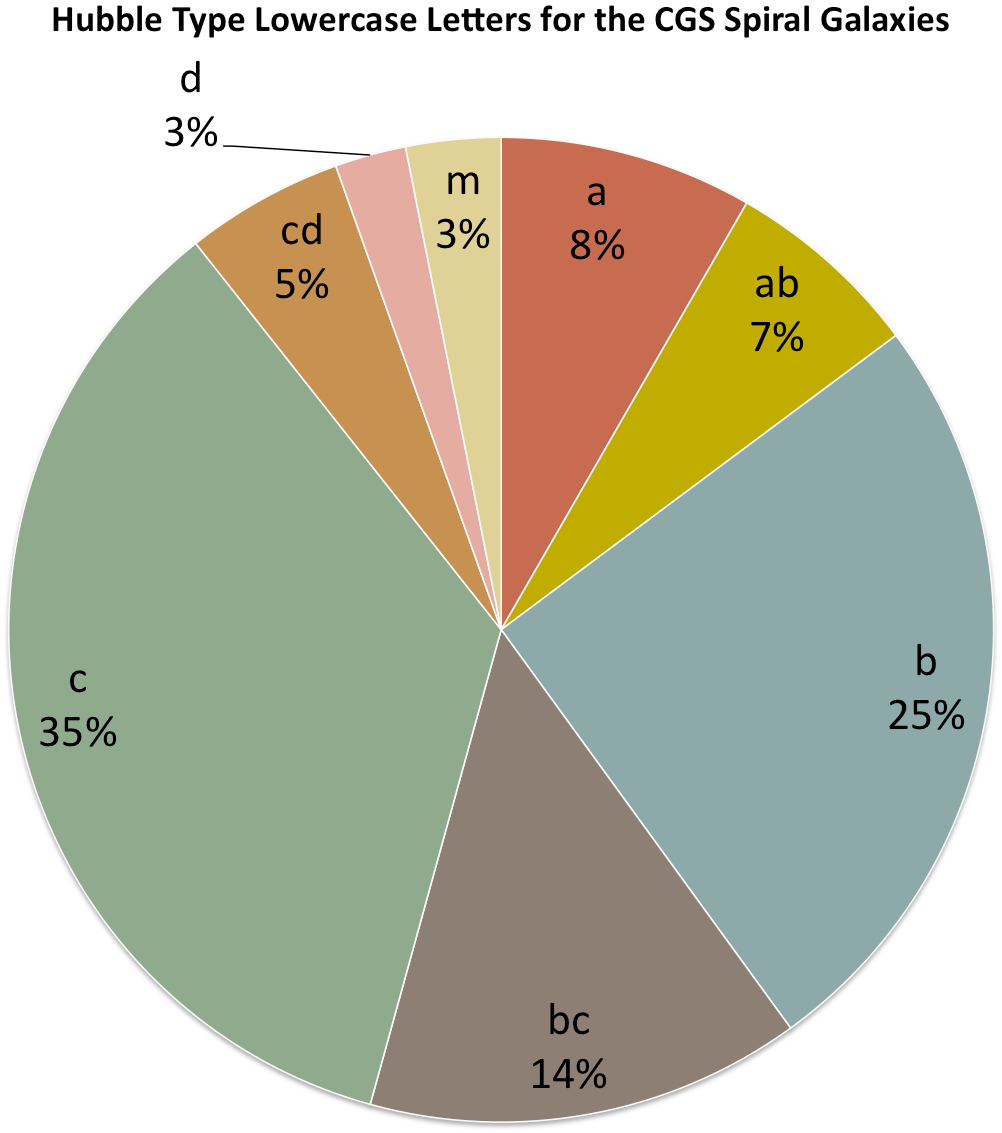}
\includegraphics[trim = 0mm 0mm 0mm 0mm, clip, width=\columnwidth]{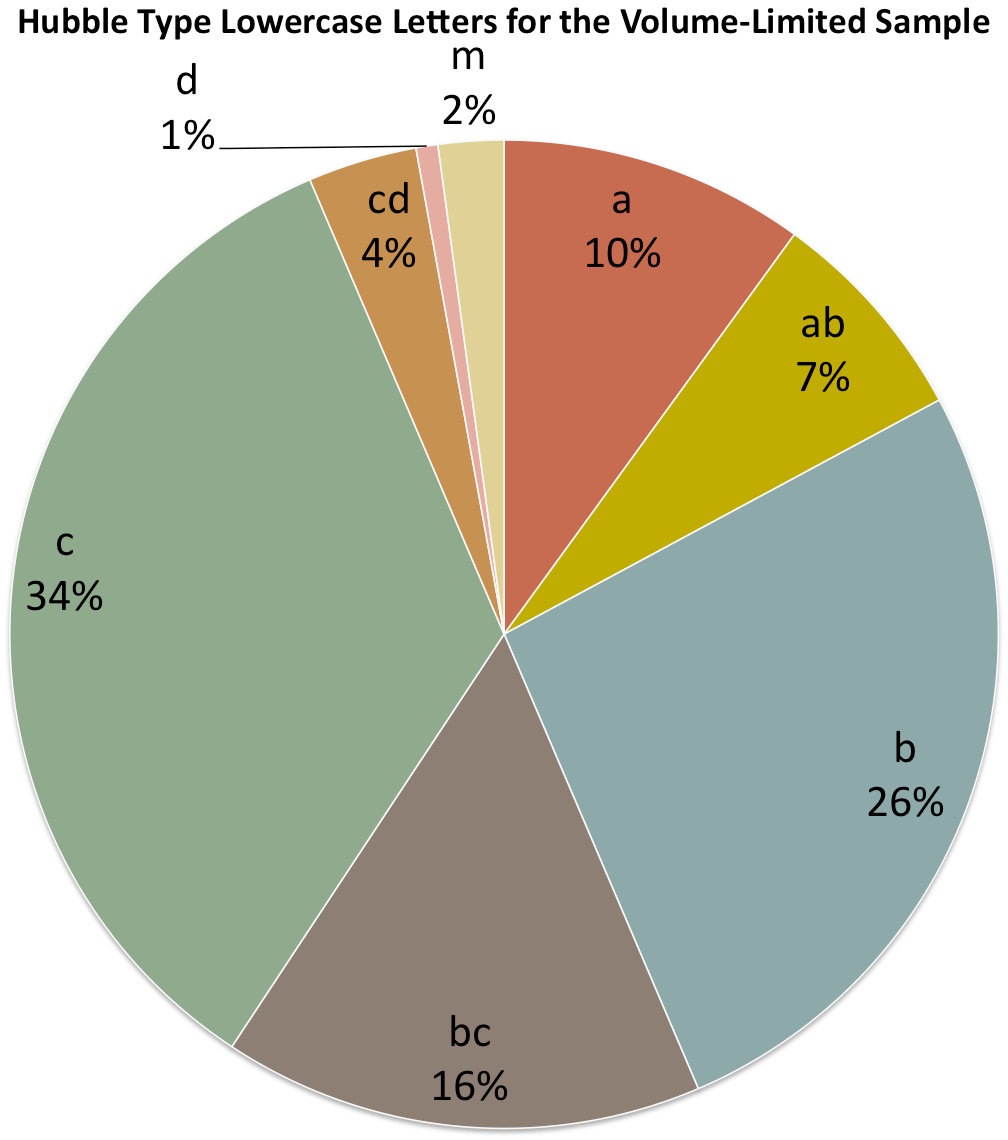}
\caption{Top: distribution of the Hubble type subdivisions (lowercase letters) for the 385 spiral galaxies contained in the CGS sample. Bottom: distribution of the Hubble type subdivisions (lowercase letters) for the 140 spiral galaxies contained in the volume-limited subsection of the CGS sample.\label{Hubble_Types}}
\end{figure}

In addition, we have determined the luminosity function
\begin{equation}
\phi(\mathfrak{M}_{\rm B}) = \frac{\partial{N}}{\partial{\mathfrak{M}_{\rm B}}},
\label{LF_Eqn}
\end{equation}
where $N$ is the number of galaxies in the sample for the volume-limited sample in terms 
of the absolute $B$-band magnitude of each galaxy and dividing by the comoving volume of the volume-limited sample (see Figure \ref{Lum_Fcn}).
The overall CGS sample has
a luminosity function very similar to that found for the much larger
Sloan Digital Sky Survey (SDSS) sample \citep{Blanton:2003}, 
indicating that it is a representative sample, in
addition to being complete or very near complete. The luminosity function
for our sample (a subset of the CGS sample) is shown in Figure \ref{Lum_Fcn}. Since we
imposed a magnitude limit of $\mathfrak{M}_{\rm B} = -19.12$ in order to maintain completeness, our
luminosity function does not extend below that limit. Above that
limit, our function seems very similar, in outline, to the luminosity
function of \citet{Blanton:2003} or the late-type galaxies from \citet{Bernardi:2013}, except for an apparent dearth 
of spiral galaxies brighter than $\mathfrak{M}_{\rm B} = -22$ in the local 
universe at distances closer than 25.4 Mpc. Additionally, our selection 
of the volume-limited sample preserved the distribution of Hubble types in 
the CGS sample, as shown in Figure \ref{Hubble_Types}.

The only notable difference between our luminosity function and that of
\citet{Blanton:2003} is found at the high-luminosity end, where our function falls
off more abruptly. The most likely explanation is that this end of the luminosity
function is dominated by a small number of very bright spiral galaxies. It is
plausible that the volume in which our sample is found is simply too small to
feature a representative number of these relatively uncommon galaxies. This
fact is obviously of some relevance to our later analysis of our black hole
mass function at the high-mass end, since we would expect very bright spirals
to have relatively large black holes.

We used imaging data taken from various sources as listed in Table \ref{Sample_Table}. Absolute magnitudes were calculated from apparent magnitudes, distance moduli, extinction factors, and $K$-corrections.  Only $B$-band absolute magnitudes were used to create a volume-limited sample.  For our local sample, the $K$-correction can be neglected.  Galactic extinction was determined from the NED Coordinate Transformation \& Galactic Extinction Calculator\footnote{\url{http://ned.ipac.caltech.edu/forms/calculator.html}}, using the extinctions values for the $B$-band from \citet{Schlafly:2011}.

\section{Pitch Angle Distribution}\label{Sect_Pitch_Dist}

\begin{figure} 
\includegraphics[trim = 27mm 0mm 0mm 8mm, clip,width=\columnwidth]{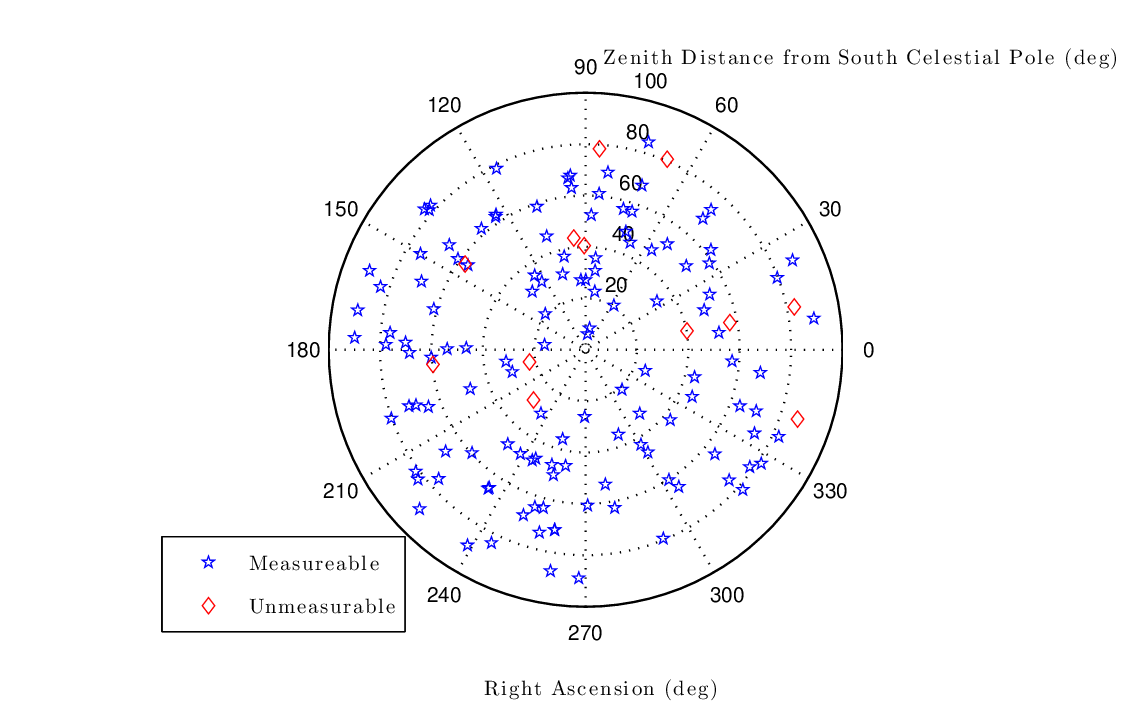}
\caption{Location of galaxies on the southern celestial hemisphere. Galaxies with measurable pitch angles are marked with \textcolor{blue}{blue} stars and galaxies with unmeasurable pitch angles are marked with \textcolor{red}{red} diamonds.\label{RA_Alt}}
\end{figure}

Pitch angle measurements were attempted for all 140 spiral galaxies in the volume-limited sample according to the method of \citet{Me:2012}. However, pitch angles were successfully measured
for only 128 of those 140 galaxies ($\approx91\%$) due to a combination of high inclination angles (10), disturbed morphology due to galaxy-galaxy interaction (1), and bright foreground star contamination (1). Overall, we achieved good coverage of the southern celestial hemisphere with our measurements (see Figure \ref{RA_Alt}) 
and the unmeasurable galaxies are randomly distributed across the southern sky. Since galaxies must first be deprojected to a face-on orientation before their pitch angle can be measured, it becomes increasingly difficult to measure galaxies where the plane of the galaxy is inclined significantly with respect to the plane of the sky. For edge-on galaxies and galaxies with extreme inclinations, it becomes impossible to recover the hidden spiral structure that is angled away from our point-of-view. Additionally, it becomes difficult to resolve spiral arms for low-surface brightness galaxies and galaxies which are too flocculent to ascertain definable spiral arms, although we avoided the former problem by
deliberately excluding the dimmest galaxies from our volume-limited sample.

\begin{figure} 
\includegraphics[trim = 0mm 0mm 0mm 0mm, clip, width=\columnwidth]{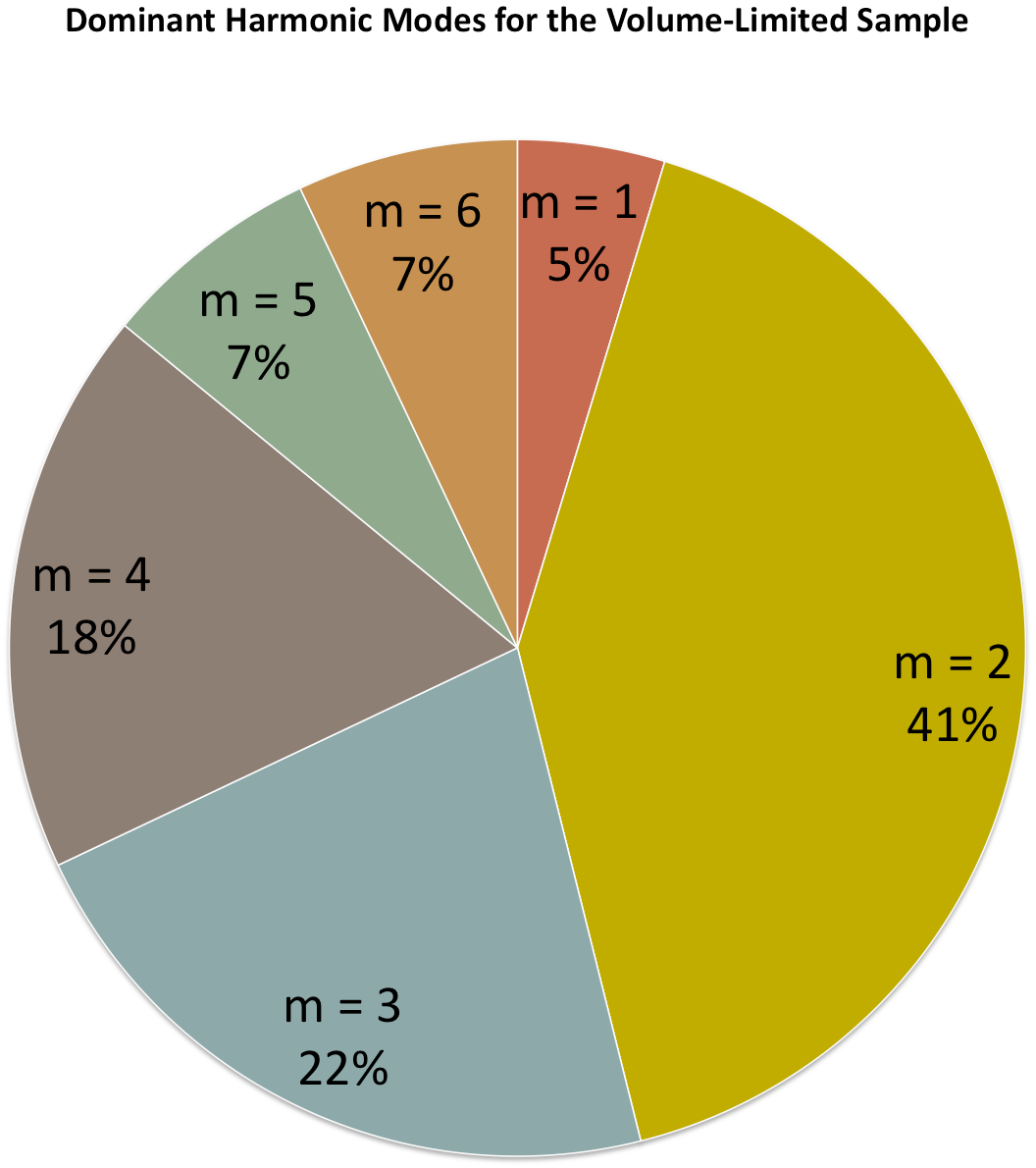}
\caption{Dominant harmonic modes ($m$) resulting from the number of spiral arms yielding the highest stability in the resulting pitch angles measured by the {\it 2DFFT} software for the sample of 128 measurable spiral galaxies from the volume-limited sample.\label{Harmonic_Modes}}
\end{figure}

All measured data for individual galaxies included in the volume-limited sample are listed in Table \ref{Sample_Table}. Approximately $55\%$ of the measurable galaxies in the volume-limited sample are observed to have positive pitch angles or clockwise chirality, with the radius of the spiral arms increasing as $\theta \to \infty$ (negative pitch angle implies counterclockwise chirality, with the radius of the spiral arms increasing as $\theta \to -\infty$). This is as expected due to the fact the sign of the pitch angle is merely a line-of-sight effect and thus, should be evenly distributed. Concerning the harmonic modes (see Figure \ref{Harmonic_Modes}), 
the $m = 2$ mode (two-armed spirals) was the most common mode ($41\%$) and the even modes constituted the majority ($66\%$). The average error on pitch angle measurements is $\pm 4.81^{\circ}$. 

NGC 5792 has the highest inclination angle amongst the galaxies with measurable pitch angles from the volume-limited sample, with an inclination angle $i = 80.44^{\circ}$\footnote{Calculated as the arccosine of the axial ratio of the minor to major axes of the galaxy.} with respect to the plane of the sky. It is important to note that this is an extreme case, and that pitch angle recovery is usually not possible for galaxies with this inclination. Only galaxies with very high resolution images, like NGC 5792, can hope to have their pitch angles determined when they are so highly inclined. Usually, a more reasonable inclination limit is $i \lessapprox 60^{\circ}$ for galaxies with average or less than average resolution. Using NGC 5792's inclination angle as a predictor of measurable inclined galaxies, the percentage of randomly inclined galaxies that would satisfy $i \leq 80.44^{\circ}$ is $\approx89\%$. This is very similar to the percentage of the volume-limited sample that we were able to measure. Of the unmeasurable 12 galaxies, 10 were too highly inclined to measure, 1 galaxy (NGC 275) was overly disturbed due to galaxy-galaxy interaction, and 1 galaxy (NGC 988) was blocked by a very bright foreground star. Due to the random nature of the unmeasurable galaxies, we still consider our volume-limited sample analysis to be statistically complete.

\begin{figure} 
\includegraphics[trim = 12mm 0mm 18mm 8mm, clip,width=\columnwidth]{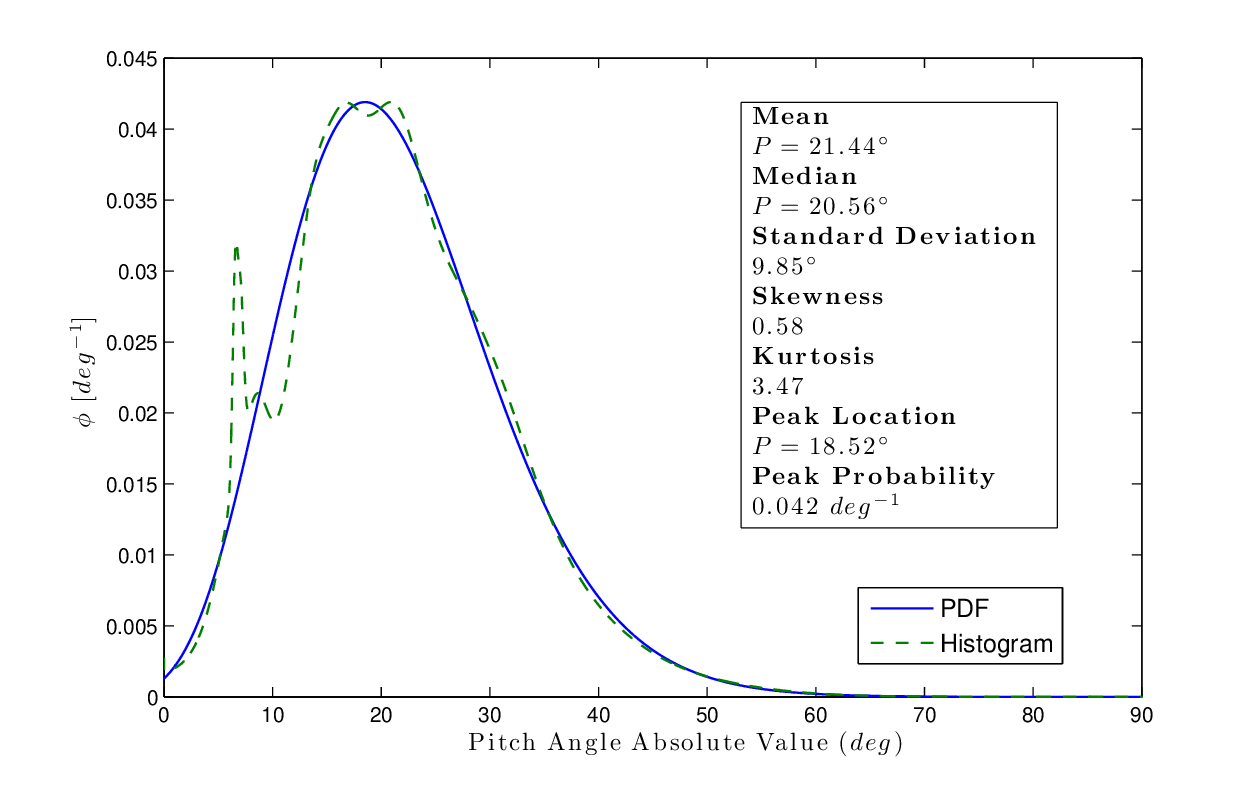}
\caption{Pitch angle distribution (dashed \textcolor[rgb]{0,0.5,0}{green} line) and a probability density function (PDF; solid \textcolor{blue}{blue} line) fit to the data. The pitch angle distribution is a ``binless" histogram that we modeled by allowing each data point to be a Gaussian, where the pitch angle absolute value is the mean and the error bar is the standard deviation. The pitch angle distribution is then the normalized sum of all the Gaussians. The resulting PDF is defined by $\mu = 21.44^{\circ}$, ${\rm median} = 20.56^{\circ}$, $\sigma = 9.85^{\circ}$, ${\rm skewness} = 0.58$, ${\rm kurtosis} = 3.47$, and a most probable pitch angle absolute value of $18.52^{\circ}$ with a probability density value of $\phi = 0.042$ deg$^{-1}$.\label{Pitch_Distribution}}
\end{figure}

In an effort to minimize the effect of Eddington bias \citep{Eddington:1913} on our data as a result of binning, we have created a nominally ``binless" pitch angle distribution from our sample of 128 galaxies, each with their individual associated errors in measurement (see Figure \ref{Pitch_Distribution}).
To do this, we constructed a routine to model each data point as a normalized Gaussian, where the pitch angle absolute value is the mean and the error bar is the standard deviation. Subsequently, the pitch angle distribution is then the normalized sum of all the Gaussians. From the resulting pitch angle distribution, we were able to compute the statistical standardized moments of a probability distribution; mean ($\mu$), variance ($\sigma^2$, quoted here by means of its square root, $\sigma$, the standard deviation), skewness, and kurtosis by analyzing the distribution with bin widths equal to the maximum resolution of our pitch angle software, $0.01^{\circ}$. Furthermore, the dimensions of the derived data array were scaled by a factor of $10^5$ to effectively smooth out the data and give the appearance of a ``binless" histogram.

In addition, we also fit a probability density function (PDF)\footnote{We use a {\it MATLAB} code called {\it pearspdf.m} to perform our PDF fittings.} to the pitch angle distribution, according to the computational results of the statistical properties of the sample ($\mu = 21.44^{\circ}$, ${\rm median} = 20.56^{\circ}$, $\sigma = 9.85^{\circ}$, ${\rm skewness} = 0.58$, and the ${\rm kurtosis} = 3.47$). From the skew-kurtotic-normal fit to the data as seen in Figure \ref{Pitch_Distribution}, it is shown that the most probable pitch angle absolute value for a galaxy is $18.52^{\circ}$, with an associated probability density value of $\phi = 0.042$ deg$^{-1}$. It is interesting to note that the most probable pitch angle is within $1.5^{\circ}$ of the pitch angle ($\left |P \right |\approx17.03^{\circ}$) of the Golden Spiral (see Appendix \ref{Golden Appendix}) and close to the pitch angle ($\left |P \right | = 22.5^{\circ} \pm 2.5^{\circ}$) of the Milky Way (see Appendix \ref{MW Appendix}). The Milky Way is a better representative of the mean pitch angle of the distribution, being only slightly greater than one degree different.

\section{Black Hole Mass Distribution}\label{Sect_Mass_Dist}

\begin{figure} 
\includegraphics[trim = 15mm 0mm 19mm 8mm, clip,width=\columnwidth]{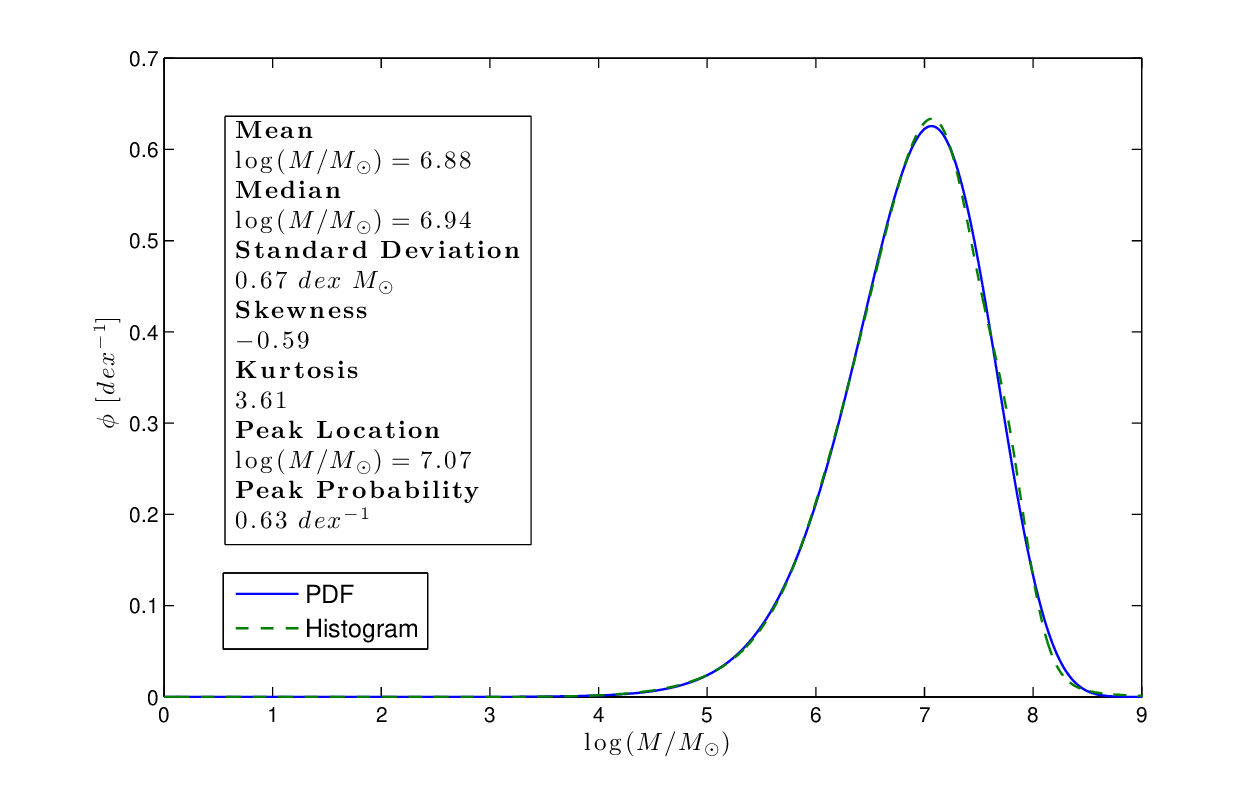}
\caption{Black hole mass distribution (dashed \textcolor[rgb]{0,0.5,0}{green} line) and a PDF (solid \textcolor{blue}{blue} line) fit to the data. The black hole mass distribution is a ``binless" histogram that we modeled by allowing each data point to be a Gaussian, where the black hole mass (converted from pitch angle measurements via Equation (\ref{M-P_Relation})) is the mean and the error bar is the standard deviation. The black hole mass distribution is then the normalized sum of all the Gaussians. The resulting PDF is defined by $\mu = 6.88$ dex $M_{\odot}$, ${\rm median} = 6.94$ dex $M_{\odot}$, $\sigma = 0.67$ dex $M_{\odot}$, ${\rm skewness} = -0.59$, ${\rm kurtosis} = 3.61$, and a most probable SMBH mass of $\log(M/M_{\odot}) = 7.07$ with a probability density value of $\phi = 0.63$ dex$^{-1}$.\label{Mass_Distribution}}
\end{figure}

The measured pitch angle values (Table \ref{Sample_Table}, Column 8) were converted to SMBH mass estimates (Table \ref{Sample_Table}, Column 11) via Equation (\ref{M-P_Relation}) with fully independent errors propagated as follows:
\begin{equation}
\delta\log(M/M_{\odot})=\sqrt{(\delta b)^2+(kP)^2[(\frac{\delta k}{k})^2+(\frac{\delta P}{P})^2]},
\label{Error_Prop}
\end{equation}
where $\delta P$ is the error associated with the pitch angle measurement. Following the procedure for creating the pitch angle distribution (see \S \ref{Sect_Pitch_Dist}), we produced a similar black hole mass distribution of the masses listed in Column 11 of Table \ref{Sample_Table} and fit a PDF to the data (see Figure \ref{Mass_Distribution}).
The resulting PDF, in terms of SMBH mass, is defined by $\mu = 6.88$ dex $M_{\odot}$, ${\rm median} = 6.94$ dex $M_{\odot}$, $\sigma = 0.67$ dex $M_{\odot}$, ${\rm skewness} = -0.59$, ${\rm kurtosis} = 3.61$, and a most probable SMBH mass of $\log(M/M_{\odot}) = 7.07$ with a probability density value of $\phi = 0.63$ dex$^{-1}$. Conversion to mass has effectively smoothed out the previous pitch angle distribution (see Figure \ref{Pitch_Distribution}), and produced a slightly more pointed (higher kurtosis) distribution. This smoothing is due to propagation of errors through Equation (\ref{M-P_Relation}), with its errors in slope and $Y$-intercept, leading to wider individual Gaussians assigned to each measurement with subsequent summation to form the black hole mass distribution in Figure \ref{Mass_Distribution}.

Nine galaxies in the sample have independently estimated SMBH masses from the literature (see Table \ref{Table_Compare}) and were included in the construction of the $M$--$P$ relation of \citet{Berrier:2013}. Rather than using these masses in our black hole mass distribution or subsequent BHMF, we chose to consistently use masses determined from the $M$--$P$ relation defined by \citet{Berrier:2013}. Our estimated masses agree with the measured masses within the listed uncertainties in all cases, as shown in Table \ref{Table_Compare}. This is not surprising given that they are included in the \citet{Berrier:2013} sample, which is defined by the directly measured masses of these galaxies (amongst others).
\begin{deluxetable}{lrcrcc} 
\tablecolumns{6}
\tablecaption{Black Hole Masses from Independent Literature Sources\label{Table_Compare}}
\tablehead{
\colhead{} & \colhead{This Work} & \colhead{} & \multicolumn{3}{c}{Literature}  \\
\cline{2-2} \cline{4-6}
\colhead{Galaxy Name} & \colhead{$\log(M/M_{\odot})$} & \colhead{} & \colhead{$\log(M/M_{\odot})$} & \colhead{Method} & \colhead{Reference}  \\
\colhead{(1)} & \colhead{(2)} & \colhead{} & \colhead{(3)} & \colhead{(4)} & \colhead{(5)}
}
\startdata
ESO 097-G013 & $6.55 \pm 0.42$ & & $6.23_{-0.08}^{+0.07}$ & 1 & 1  \\
Milky Way & $6.82 \pm 0.30$ & & $6.63 \pm 0.04$ & 2 & 2  \\
NGC 253 & $6.92 \pm 0.54$ & & $\approx6.94$ & 1 & 3  \\
NGC 1068 & $6.93 \pm 0.37$ & & $6.88 \pm 0.02$ & 1 & 4  \\
NGC 1300 & $7.42 \pm 0.23$ & & $7.80 \pm 0.29$ & 3 & 5  \\
NGC 1353 & $7.36 \pm 0.25$ & & $6.64 \pm 0.71$ & 4 & 6  \\
NGC 1357 & $7.21 \pm 0.31$ & & $7.19 \pm 0.71$ & 4 & 6  \\
NGC 3621 & $7.43 \pm 0.21$ & & $\gtrapprox 3.64$ & 5 & 7  \\
NGC 7582 & $7.30 \pm 0.51$ & & $7.75_{-0.18}^{+0.17}$ & 3 & 8  \\
\enddata
\tablecomments{Columns: (1) Galaxy name (in order of increasing R.A.). (2) SMBH mass in $\log(M/M_{\odot})$, converted from the pitch angle by Equation (\ref{M-P_Relation}). (3) SMBH mass in $\log(M/M_{\odot})$, as listed by independent literature sources (when applicable, masses have been adjusted to conform with our defined cosmology). (4) SMBH mass estimation method used by independent literature source. (5) Literature source of SMBH mass.
Method:
(1) ${\rm H_2O}$ maser;
(2) stellar orbits;
(3) gas dynamics;
(4) $M$--$\sigma$ relation;
(5) Eddington limit.}
\tablerefs{
(1) \citet{Greenhill:2003};
(2) \citet{Levine:2006};
(3) \citet{Rodriguez-Rico:2006};
(4) \citet{Lodato:Bertin:2003};
(5) \citet{Atkinson:2005};
(6) \citet{Ferrarese:2002};
(7) \citet{Satyapal:2007};
(8) \citet{Wold:2006}.}
\end{deluxetable}

It is also worth noting that half a dozen galaxies included in our volume-limited sample harbor nuclear star clusters (NSC) with well-determined masses \citep{Erwin:Gadotti:2012}. The existence of a NSC in a galaxy does not rule out the coexistence of a SMBH and vice versa. For instance, the Milky Way and M31 have been shown to both contain a NSC and a SMBH, all with well-determined masses \citep{Erwin:Gadotti:2012}. It has been shown that NSCs and SMBHs do not follow the same host-galaxy correlations; SMBH mass correlates with the stellar mass of the bulge component of galaxies, while NSC mass correlates much better with the total galaxy stellar mass \citep{Erwin:Gadotti:2012}. Because of this, our implied SMBH masses for these seven galaxies is not equivalent to the known masses of their NSCs, their only known central massive objects (see Table \ref{NSCs}). By comparing the central massive objects in Table \ref{NSCs}, it can be seen that the average NSC mass is higher than the average SMBH mass for this sample; $\log(M/M_{\odot}) = 7.55 \pm 0.16$ and $\log(M/M_{\odot}) = 7.04_{-0.25}^{+0.28}$, respectively.
\begin{deluxetable}{lrrrrc} 
\tablecolumns{6}
\tablecaption{Galaxies with Well-determined NSC Masses\label{NSCs}}
\tablehead{
\colhead{} & \colhead{SMBHs} & \colhead{} & \colhead{NSCs}  \\
\cline{2-2} \cline{4-4}
\colhead{Galaxy Name} & \colhead{$\log(M/M_{\odot})$} & \colhead{} & \colhead{$\log(M/M_{\odot})$} & \colhead{} & \colhead{Reference}  \\
\colhead{(1)} & \colhead{(2)} & \colhead{} & \colhead{(3)} & \colhead{} & \colhead{(4)}
}
\startdata
Milky Way & $6.82 \pm 0.30$ & & $7.48_{-0.30}^{+0.18}$ & & 1  \\
NGC 1325 & $7.35 \pm 0.21$ & & $7.12 \pm 0.30$ & & 2  \\
NGC 1385 & $5.99 \pm 0.49$ & & $6.30 \pm 0.30$ & & 2  \\
NGC 3621 & $7.43 \pm 0.21$ & & $\approx7.01$ & & 3  \\
NGC 4030 & $6.75 \pm 0.44$ & & $8.05 \pm 0.30$ & &2  \\
NGC 7418 & $6.58 \pm 0.59$ & & $7.75 \pm 0.19$ & & 4  \\
\enddata
\tablecomments{Columns: (1) Galaxy name. (2) SMBH mass in $\log(M/M_{\odot})$, converted from the pitch angle by Equation (\ref{M-P_Relation}). (3) NSC mass in $\log(M/M_{\odot})$. (4) Source of NSC measurement.}
\tablerefs{
(1) \citet{Launhardt:2002};
(2) \citet{Rossa:2006};
(3) \citet{Barth:2009};
(4) \citet{Walcher:2005}.}
\end{deluxetable}

Ultimately, Figure \ref{Mass_Distribution} provides us with a look at a simple 1:1 conversion from pitch angle to SMBH mass via the $M$--$P$ relation. Since this only applies to the 128 measurable galaxies (out of the total volume-limited sample of 140 galaxies), it offers the most direct look at the distribution of SMBH masses in the Local Universe. The subsequent section will extend the results into the complete BHMF via extrapolation to the full 140 member volume-limited sample and full treatment of sampling from probability distributions.

\section{Black Hole Mass Function for Local Spiral Galaxies}\label{Sect_BHMF}

The pitch angle function $\phi(P)$ is defined as
\begin{equation}
\phi(P) = \frac{\partial{N}}{\partial{P}},
\label{Pitch_Angle_Function}
\end{equation}
where $\phi(P)dP$ is defined to be the number of galaxies that have pitch angles between $P$ and $P + dP$. That should be $\frac{\partial{N}}{\partial{P}}dP$ because
\begin{equation}
N = \int_{0}^{\pi}\frac{\partial{n}}{\partial{P}}dP
\label{Total_Number}
\end{equation}
is the total number of galaxies in the sample. Then the BHMF is
\begin{equation}
\phi(M) = \frac{\partial{N}}{\partial{M}} = \frac{\partial{N}}{\partial{P}}\frac{\partial{P}}{\partial{M}} = \phi(P)\frac{\partial{P}}{\partial{M}}.
\label{Mass_Function}
\end{equation}
Therefore, by taking the derivative of Equation (\ref{M-P_Relation}) we find
\begin{equation}
\frac{1}{M\ln(10)} = -(k \pm \delta k)\frac{\partial{P}}{\partial{M}}
\label{M-P_Relation_Derivative}
\end{equation}
or
\begin{equation}
\frac{\partial{P}}{\partial{M}} = -\frac{1}{M\ln(10)(k \pm \delta k)}.
\label{Rearranged}
\end{equation}
Therefore,
\begin{equation}
\phi(M) = -\frac{\phi(P)}{M\ln(10)(k \pm \delta k)}.
\label{Mass Solution}
\end{equation}
Alternatively, we could calculate
\begin{equation}
\phi(\log M) = \frac{\partial{N}}{\partial{\log M}} = \frac{\partial{N}}{\partial{P}}\frac{\partial{P}}{\partial{\log M}} = -\frac{\phi(P)}{k \pm \delta k}.
\label{LogM_Function}
\end{equation}

Through the implementation of Equation (\ref{LogM_Function}) and dividing by the comoving volume of the volume-limited sample ($V_{\rm C} = 3.37$ $\times$ $10^4$ $h_{67.77}^{-3}$ Mpc$^3$), the pitch angle PDF in Figure \ref{Pitch_Distribution} can be transformed into a BHMF.
Using the probabilities established by the PDF in Figure \ref{Pitch_Distribution}, we can predict probable masses for the remaining 12 unmeasurable galaxies in the volume-limited sample, in order to extrapolate the BHMF and related parameters for the full sample size. The evaluation of BHMF with the summation of all SMBH masses and total densities for both the measurable sample of 128 SMBHs and the extrapolated full volume-limited sample of 140 SMBHs are listed in Table \ref{BHMF_Evaluation}.
\begin{deluxetable}{lcccccc} 
\tablecolumns{5}
\tablecaption{Black Hole Mass Function Evaluation\label{BHMF_Evaluation}}
\tablehead{
\colhead{N} & \colhead{$M_{\rm Total}$} & \colhead{$\rho$} & \colhead{$\Omega_{\rm BH}$} & \colhead{$\Omega_{\rm BH}/\omega_{\rm b}$}  \\
\colhead{} & \colhead{($10^9$ $M_{\odot}$)} & \colhead{($10^4$ $h_{67.77}^3$ $M_{\odot}$ Mpc$^{-3}$)} & \colhead{($10^{-7}$ $h_{67.77}$)} & \colhead{($h_{67.77}^3$ $\permil$)}  \\
\colhead{(1)} & \colhead{(2)} & \colhead{(3)} & \colhead{(4)} & \colhead{(5)}
}
\startdata
128 & $1.75_{-0.85}^{+2.05}$ & $5.17_{-2.52}^{+6.07}$ & $4.06_{-1.98}^{+4.76}$ & $0.018_{-0.009}^{+0.022}$  \\
140 & $1.87_{-0.92}^{+2.21}$ & $5.54_{-2.73}^{+6.55}$ & $4.35_{-2.15}^{+5.14}$ & $0.020_{-0.010}^{+0.023}$  \\
\enddata
\tablecomments{Columns: (1) Number of galaxies (measurable 128 SMBHs or an extrapolation for the full volume-limited sample of 140 SMBHs). (2) Total mass from the summation of all the SMBHs in units of $10^9$ $M_{\odot}$. (3) Density (Column (2) divided by $3.37$ $\times$ $10^4$ $h_{67.77}^{-3}$ Mpc$^3$) of SMBHs in units of $10^4$ $h_{67.77}^3$ $M_{\odot}$ Mpc$^{-3}$. (4) Cosmological SMBH mass density for spiral galaxies ($\Omega_{\rm BH} = \rho/\rho_0$), assuming $\rho_{0} \equiv 3H_{0}^2/8\pi G = 1.274$ $\times$ $10^{11}$ $M_{\odot}$ Mpc$^{-3}$ when $H_0 = 67.77$ km s$^{-1}$ Mpc$^{-3}$. (5) Fraction of the universal baryonic inventory locked up in SMBHs residing in spiral galaxies ($\Omega_{\rm BH}/\omega_{\rm b}$).}
\end{deluxetable}

\begin{figure} 
\includegraphics[trim = 5mm 5mm 0mm 0mm, clip, width=\columnwidth]{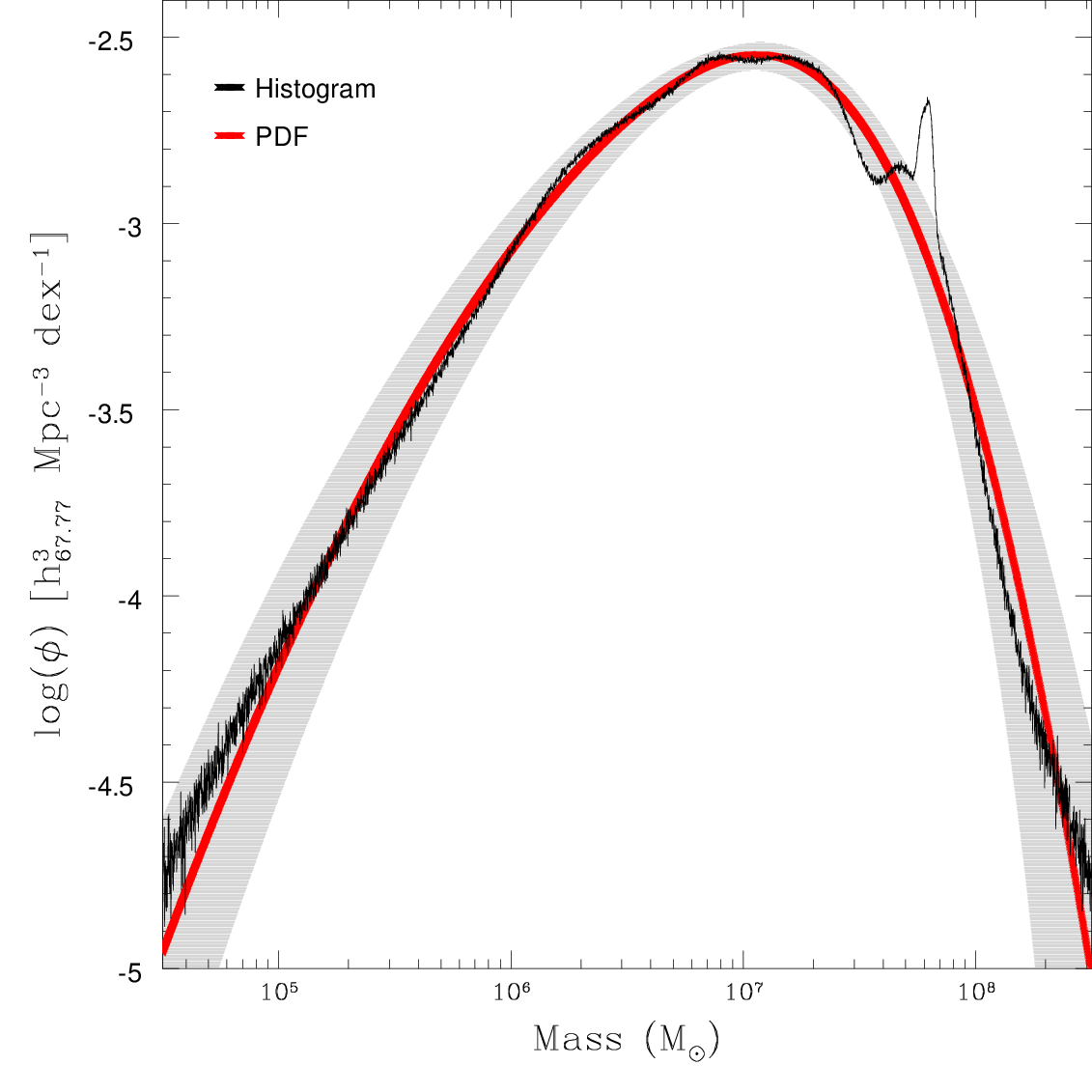}
\includegraphics[trim = 9mm 5mm 0mm 0mm, clip, width=\columnwidth]{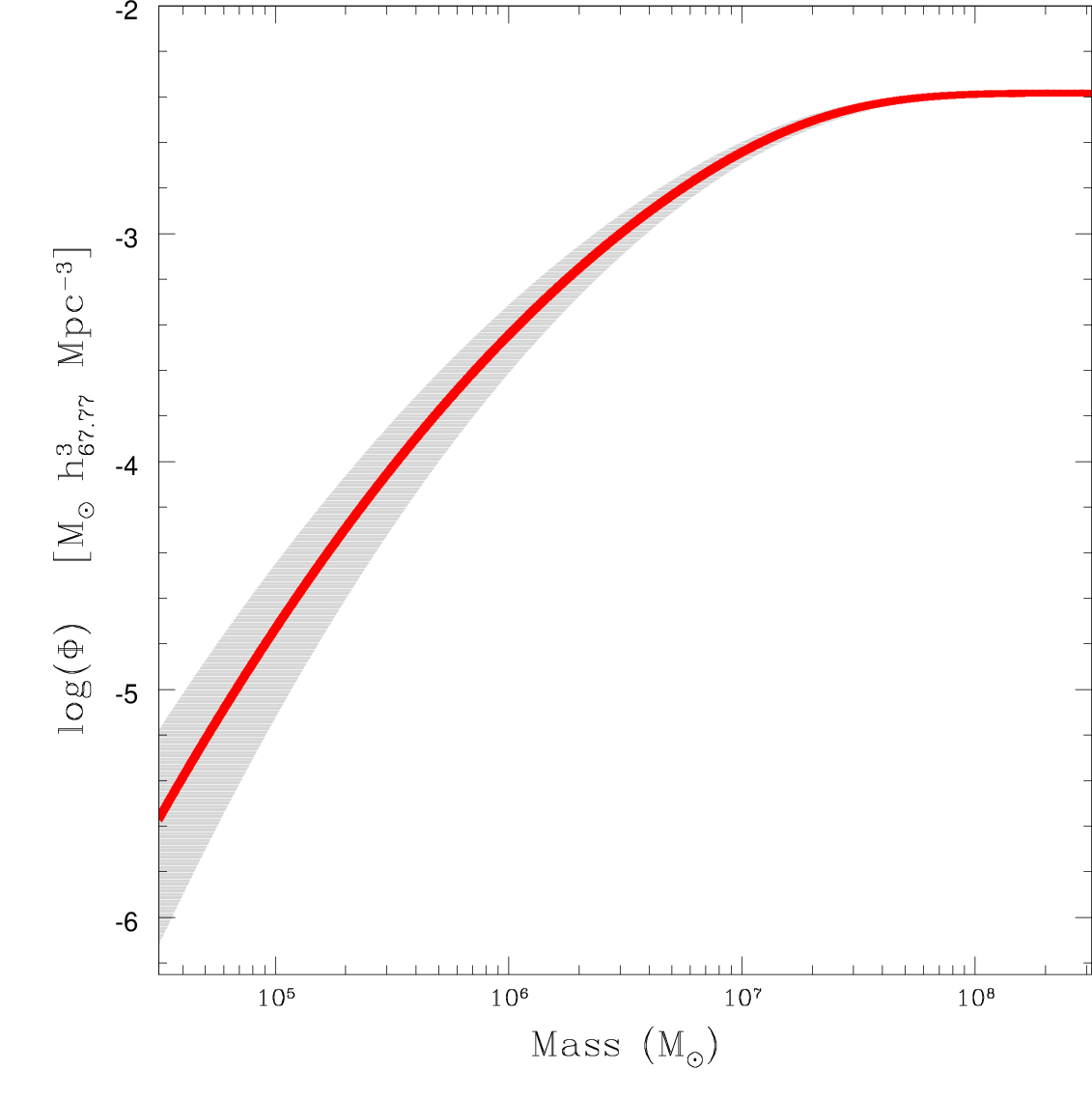}
\caption{Top: MCMC sampling of the late-type BHMF (rough {\bf black} line) with the best fit model PDF (solid \textcolor{red}{red} line) surrounded by a $\pm 1\sigma$ error region (\textcolor{gray}{gray} shading). When integrated, the area under the curve yields the number density for the entire volume-limited sample, 4.15 $\times$ $10^{-3}$ $h_{67.77}^3$ Mpc$^{-3}$. The plotted data for the top panel is listed for convenience in Table \ref{Values}. Bottom: MCMC sampling of the late-type BHMF with the best fit model CDF (solid \textcolor{red}{red} line) surrounded by a $\pm 1\sigma$ error region (\textcolor{gray}{gray} shading). The CDF visually depicts the integration of the above PDF in the top panel from $M = 0$ until any desired reference point. Here, $\Phi$ is used to indicate an integrated probability, elsewhere $\phi$ is used to indicate a probability density. The upper asymptote approaches the number density for the entire volume-limited sample, 4.15 $\times$ $10^{-3}$ $h_{67.77}^3$ Mpc$^{-3}$.\label{Monte_Carlo}}
\end{figure}

\begin{figure} 
\includegraphics[trim = 12mm 5mm 0mm 0mm, clip, width=\columnwidth]{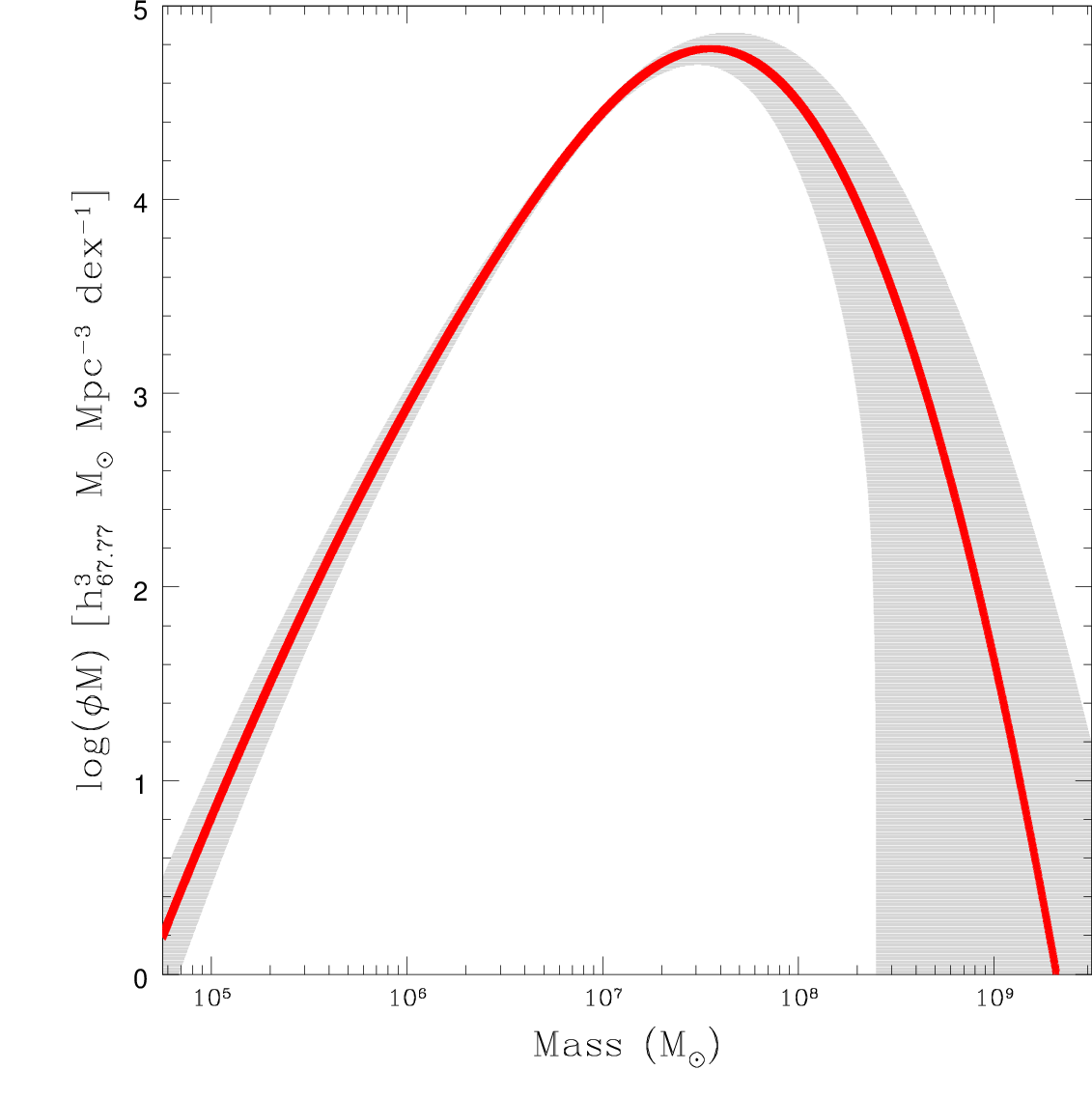}
\caption{Contribution by the SMBH mass to the cosmic SMBH mass density (solid \textcolor{red}{red} line) surrounded by a $\pm 1\sigma$ error region (\textcolor{gray}{gray} shading). This plot is proportional to Figure \ref{Monte_Carlo} (top), in that this is the product of the BHMF and the SMBH mass ($\phi M$). When integrated, the area under the curve for this plot yields the SMBH mass density, $\rho = 5.54_{-2.73}^{+6.55}$ $\times$ $10^4$ $h_{67.77}^3$ $M_{\odot}$ Mpc$^{-3}$.\label{Mass_Density}}
\end{figure}

In order to determine errors on the calculated late-type BHMF, we ran a Markov Chain Monte Carlo (MCMC) sampling\footnote{We perform the sampling with a modified {\it C} version of the original {\it Python} implementation \citep{MCMC} of an affine-invariant ensemble sampler \citep{Hou:2012} using an ensemble of 1000 walkers.} of the late-type BHMF. The sampling consisted of $10^5$ realizations for each of the 128 measured galaxies, with pitch angles randomly generated from the data with a Gaussian Distribution within $5\sigma$ of each measured pitch angle value. In addition, the fit to the $M$--$P$ relation (Equation (\ref{M-P_Relation})) was also allowed to vary based on the intrinsic errors in slope and $Y$-intercept, which again assumes a Gaussian distribution around the fiducial values. Ultimately, SMBHs are determined from pitch angle values using both the fiducial and randomly adjusted fit. Comparison between the two samples allowed us to represent the fit to the late-type BHMF with error regions. We display the results both as a PDF and a cumulative density function (CDF) fit to the data (see Figure \ref{Monte_Carlo}).
The plotted data for Figure \ref{Monte_Carlo} (top) is listed for convenience in Table \ref{Values}. The location of the peak and its value for the MCMC PDF are $\log(M/M_{\odot}) = 7.07_{-0.09}^{+0.09}$ and $\phi = 2.84_{-0.23}^{+0.26}$ $\times$ $10^{-3}$ $h_{67.77}^3$ Mpc$^{-3}$ dex$^{-1}$, respectively. Additionally, we provide a proportional plot for Figure \ref{Monte_Carlo} (top), in terms of the product of the BHMF probability density and the SMBH mass ($\phi M$), showing the contribution by the SMBH mass to the cosmic SMBH mass density (see Figure \ref{Mass_Density}).
\begin{deluxetable}{lr} 
\tablecolumns{2}
\tablecaption{BHMF MCMC PDF Values\label{Values}}
\tablehead{
\colhead{$\log(M/M_{\odot})$} & \colhead{$\phi$ ($10^{-4}$ $h_{67.77}^3$ Mpc$^{-3}$ dex$^{-1}$)}  \\
\colhead{(1)} & \colhead{(2)}
}
\startdata
5.00 & $0.65_{-0.36}^{+0.52}$  \\
5.25 & $1.38_{-0.67}^{+0.87}$  \\
5.50 & $2.74_{-1.14}^{+1.34}$  \\
5.75 & $5.01_{-1.72}^{+1.88}$  \\
6.00 & $8.46_{-2.33}^{+2.39}$  \\
6.25 & $13.19_{-2.82}^{+2.76}$  \\
6.50 & $18.92_{-3.01}^{+2.83}$  \\
6.75 & $24.67_{-2.83}^{+2.47}$  \\
7.00 & $28.23_{-2.58}^{+2.18}$  \\
7.25 & $26.53_{-3.02}^{+2.62}$  \\
7.50 & $18.90_{-3.21}^{+3.06}$  \\
7.75 & $9.49_{-2.91}^{+3.22}$  \\
8.00 & $3.19_{-1.77}^{+2.34}$  \\
8.25 & $0.69_{-0.58}^{+1.07}$  \\
8.50 & $0.09_{-0.09}^{+0.32}$  \\
\enddata
\tablecomments{Columns: (1) SMBH mass listed as $\log(M/M_{\odot})$ in $0.25$ dex intervals. (2) BHMF number density values from the resulting PDF fit to the MCMC sampling at the given mass in units of $10^{-4}$ $h_{67.77}^3$ Mpc$^{-3}$ dex$^{-1}$.}
\end{deluxetable}

\begin{figure}  
\includegraphics[trim = 0mm 1mm 0mm 0mm, clip, width=\columnwidth]{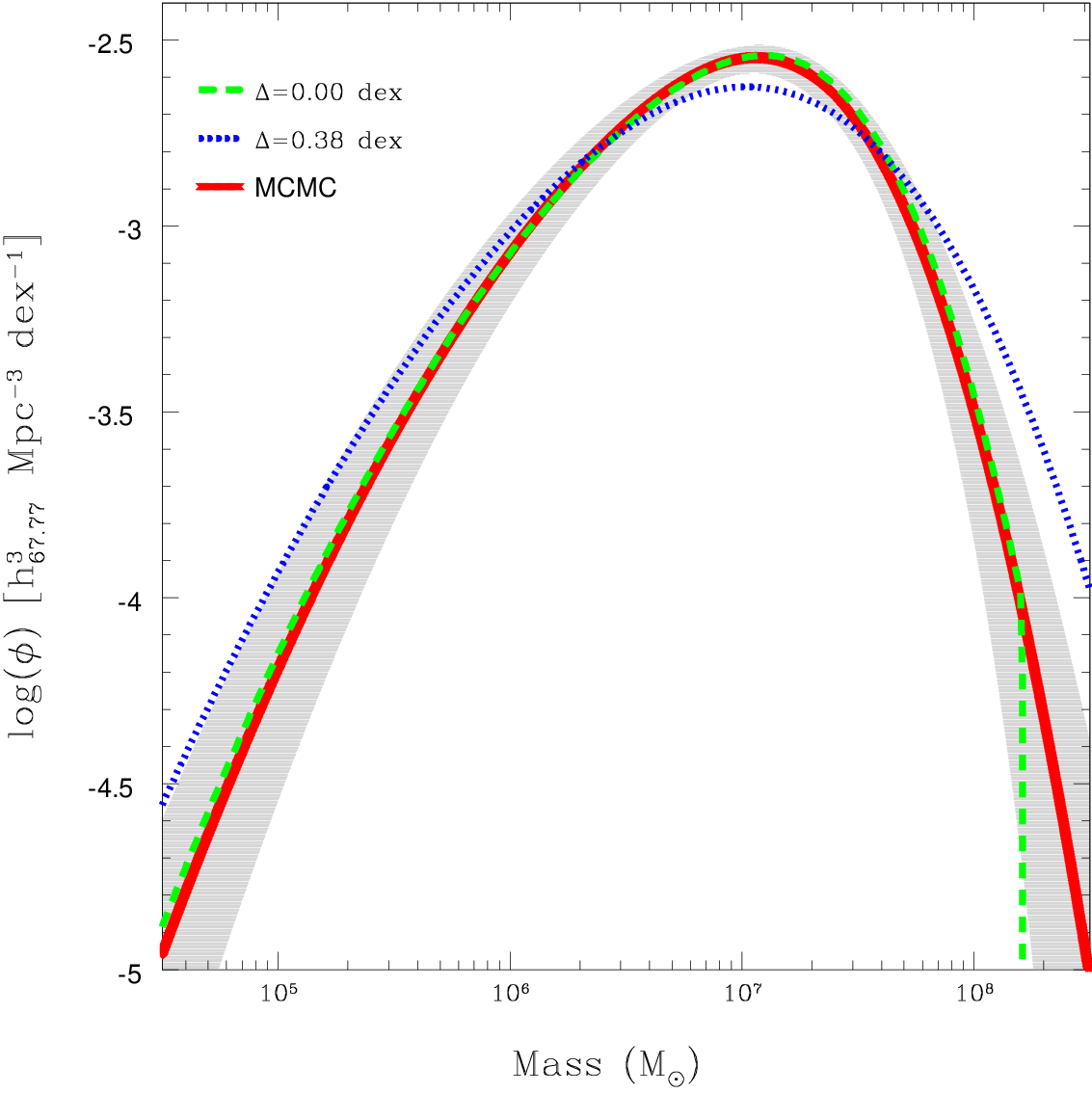}
\caption{Comparison of the BHMFs generated through MCMC sampling (solid \textcolor{red}{red} line with \textcolor{gray}{gray} shading) and through convolution of the probabilities associated with zero (dashed \textcolor{green}{green} line) and 0.38 dex (dotted \textcolor{blue}{blue} line) intrinsic dispersions.\label{Convolved_BHMF}}
\end{figure}

Since the role played by the intrinsic error in the $M$--$P$ relation is of particular
interest, we also adopted the procedure described in \citep{Marconi:2004} (see Equation (3) of
that paper and the surrounding discussion) which convolves the distribution function of
(in our case) pitch angles in our sample with a Gaussian distribution representing the
intrinsic scatter of the $M$--$P$ relation. Since the true intrinsic scatter of this relation
is unknown, we simply used the maximum dispersion of 0.38 dex found in \citep{Berrier:2013}. 
In reality,
the intrinsic dispersion is presumably somewhat less than this, since at least some of
the scatter found in that paper must be due to measurement errors (of both pitch angle
and black hole mass). The result of this calculation is a mass function that is broader
than that discussed previously because we allow for the possibility that some galaxies
are misplaced due to an intrinsic uncertainty in translating from a pitch angle measurement
to a black hole mass. The natural result is to broaden the mass function, as compared to
one with no intrinsic dispersion assumed. In Figure \ref{Convolved_BHMF}, 
we see that on the low-mass side
this calculation agrees very well with the outer $1\sigma$ error region from the MCMC calculation. This is not surprising since both the convolution technique and the
MCMC calculation account for intrinsic dispersion as well as measurement error in
pitch angle. It is evident that the zero intrinsic dispersion BHMF is very similar to the MCMC BHMF, except for the abrupt stop of the zero intrinsic dispersion BHMF at $\log(M/M_{\odot}) = 8.21$, due to the $Y$-intercept of the $M$--$P$ relation. On the high-mass side, the convolution technique actually broadens the mass
function even more and this is significant, as we will see in the next section, in view
of comparisons to be made with mass function derived from other techniques.

\section{Discussion}\label{Sect_Discussion}

\begin{figure} 
\includegraphics[trim = 0mm 1mm 0mm 0mm, clip, width=\columnwidth]{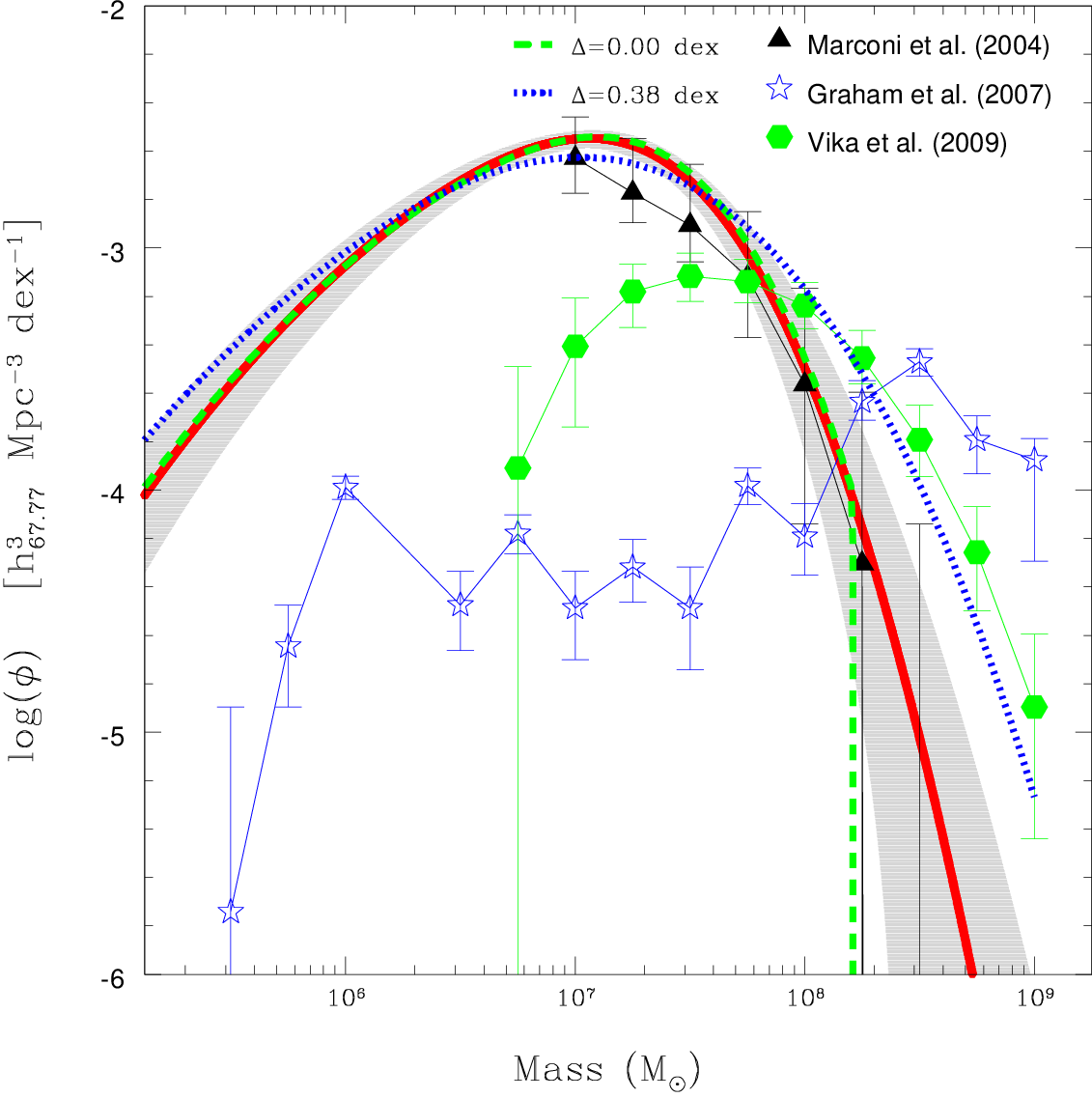}
\caption{Comparison between our determination of the BHMF for late-type galaxies with our MCMC fit in \textcolor{red}{red} with a \textcolor{gray}{gray} 
shaded $\pm1\sigma$ error region, zero intrinsic dispersion (dashed \textcolor{green}{green} line), and 0.38 dex intrinsic dispersion (dotted \textcolor{blue}{blue} line); with those of \citet{Marconi:2004}, depicted by {\bf black} triangles 
(a late-type BHMF is not provided in \citet{Marconi:2004}, we have merely subtracted their 
early-type function from their all-type function); \citet{Graham:2007}, depicted 
by \textcolor{blue}{blue} stars; and \citet{Vika:2009}, depicted by \textcolor{green}{green} 
hexagons. The BHMF of \citet{Vika:2009} is derived using a relationship between SMBH mass and 
the luminosity of the host galaxy spheroid, applied to a dust-corrected sample of 312 late-type 
galaxies from the Millennium Galaxy Catalogue in the redshift range $0.013 \leq z \leq 0.18$. 
The peak of our BHMF is located at $\log(M/M_{\odot}) = 7.06$, whereas theirs is located 
at $\log(M/M_{\odot}) = 7.50$. However, \citet{Vika:2009} consider BHMF data for $\log(M/M_{\odot}) < 7.67$ 
to be unreliable because it is derived from galaxies with $\mathfrak{M}_{\rm B} > -18$, according 
to their relationship. Note that our entire sample consists of galaxies with $\mathfrak{M}_{\rm B} \leq -19.12$.\label{Late-Types}}
\end{figure}

Compared to the attention focused on
the early-type mass function, there have been notably less efforts to estimate the local BHMF for
spiral or late-type galaxies.\footnote{One must say a word, at this point, on the question of
whether lenticular galaxies (Hubble Type {\it S0}) should be
included with early types or late types.
Generally, in the BHMF literature they are counted as early types.
This is understandable, since it is probably more straightforward
to apply bulge-related correlations, such as 
$M--\sigma$ to them than it is for spiral-armed galaxies.
Since they have no visible spiral arms, they are clearly unsuitable for our
method. 
We obviously do not include lenticulars in our mass function. We also
do not include edge-on galaxies but this should surely be randomly
selected and our luminosity function does not show any sign of a
systematic loss.} Even studies that produce a BHMF for
all types of galaxies will often use a different 
procedure for producing the late-type portion of it. An example
is that of \citet{Marconi:2004}, which uses a velocity dispersion relation
for early-type galaxies in the {\it SDSS} based on 
actual measurements of $\sigma$. They include a
BHMF for all galaxy types as well, from which one can deduce their
late-type BHMF (see Figure \ref{Late-Types}). 
Their data for late-type galaxies is based on a
velocity dispersion function given by \citet{Sheth:2003}, who appear
to define late types as being spiral galaxies, as we do, including
lenticulars with the early types. They make use of the Tully--Fisher
relation \citep{Tully:Fisher:1977} to convert the luminosity function
of late types in the SDSS into a function of the circular velocities
of these galaxies (meaning that the typical rotational velocity of each
of their galactic disks) and then use an observed and expected correlation
between these circular velocities and the velocity dispersion ($\sigma$)
of their bulges to obtain a velocity dispersion relation for late types.
\citet{Marconi:2004} then invoke the $M$--$\sigma$ relation to convert this into
a BHMF. It is worth noting the number of steps involved in this process
and the fact that the final product does not incorporate any data from the
SDSS beyond the luminosity function of the galaxies involved. 
It is obviously encouraging that the results of \citet{Marconi:2004} agree so well
with our mass function (assuming zero intrinsic dispersion in the $M$--$P$ relation)
at the high-mass end (see Figure \ref{Late-Types}). We cannot compare at the low-mass end, where \citet{Marconi:2004}
do not provide any data.

In Figure \ref{Late-Types}, we also compare to the work of \citet{Graham:2007}, which is based
on measurements of the S\'ersic index of the galactic bulge.
As can be seen in Figure \ref{Late-Types}, it is difficult to interpret the data
of \citet{Graham:2007} for late-type galaxies and this may be due to the increased difficulty
in extracting S\'ersic index values for this type of galaxy, where one
must disentangle multiple galactic components (disk and often bar) in
order to obtain the S\'ersic index (A. Graham 2012, private communication). Our
numbers agree far better with those found in \citet{Marconi:2004}. 

An example of more recent work with which we can compare
is the late-type BHMF presented in \citet{Vika:2009}, which is based
upon measurements of bulge luminosities in late-type galaxies in the
SDSS. Figure \ref{Late-Types} also compares our BHMF with theirs. 
At the very high-mass end our mass function, allowing for the
intrinsic dispersion of the $M$--$P$ relation, it comes quite close to the
mass function of \citet{Vika:2009}. At the middle and low-mass end,
in contrast, their mass function is far below what we find.

\citet{Vika:2009} use the SDSS while our sample is
based upon the selection in the CGS, which is considerably more local. Our
most distant galaxy has (in our cosmology) a redshift of 0.00572. Their
nearest galaxy has a redshift of 0.013 and their most distant
is close to $z = 0.18$. They have 312 objects in their late-type
sample, we have roughly half that. However, the volume of their sample is
considerably larger than ours ($\approx41$ times), so we would expect more late-types
in theirs if they were sampling the same types of galaxies as ours.
Given that their sample\footnote{\citet{Graham:2007} uses the same parent sample, the {\it Millennium Galaxy Catalogue}, but uses only 230 late-type galaxies.} is more distant, it seems likely that they
are missing the dimmer galaxies, which would tend to explain the relative
scarcity of smaller black holes in their BHMF. 
On the other hand, their much larger sample volume makes it more likely that
they have observed the brighter spirals that may be missing from our sample,
based on the luminosity function comparison shown in Figure \ref{Lum_Fcn}.

Comparing the 
\citet{Vika:2009} late-type mass function with 
ours (from Figure \ref{Late-Types}), we are struck overall by the generally good
agreement we find. Although there is some disagreement between \citet{Marconi:2004} and \citet{Vika:2009}
at the high-mass end, the comparison with our results sheds some light on
a possible reason. We agree
very closely with \citet{Marconi:2004} when assuming no error in the $M$--$P$ relation, and are
quite close to \citet{Vika:2009} when we assume that all of the scatter in the $M$--$P$ relation
is due to an intrinsic dispersion in that correlation. Since presumably at least
some of that scatter is merely due to measurement error in either $M$ or $P$, it
is natural to expect that the true SMBH at the high-mass end falls somewhere between
the curves given by \citet{Marconi:2004} and \citet{Vika:2009}. It should be kept in mind that the
evidence of a deficit in very bright spirals in our volume-limited sample does
lend credence to the view that the final result may be close to the line given by
\citet{Vika:2009} at the very high-mass end. However, in addition, the fact that the high-mass
end of the black hole spectrum is dominated by a relatively small number
of large objects is one explanation of why a certain level of disagreement
is not altogether unexpected in this regime. In short, it looks as if
\citet{Marconi:2004} and \citet{Vika:2009} each fall at opposite
limits of our error bars in this
regime, which suggests that none of the three results are in severe conflict
with each other.


In the low-mass end, there is much less data with which we can compare.
\citet{Vika:2009} disagree strongly with us on the low-mass end. Their
data is based
on a sample drawn from the SDSS, which covers a much larger volume of space 
than our sample, which is based on the most local part of the CGS. In spite of this, \citet{Vika:2009} have only twice as many late-type
 galaxies in their total sample as we do. It 
seems likely that \citet{Vika:2009} miss many galaxies because they are too dim 
to be easily observed at the greater range of their sample. This could 
explain the fact that we see far more smaller black holes than they do.
Therefore, we conclude that we are not yet in a position to compare with
any similarly complete surveys in this particular regime. The good
agreement we enjoy with other results at the high-mass end obviously
gives grounds for optimism on the low-mass end. We have made a considerable
effort to provide a complete local sample precisely because of our interest
in producing reliable data on the low-mass end of the black hole spectrum.
Obviously, since we have a luminosity cutoff, we must accept that we could be
missing black holes at the low-mass end, black holes which would reside
in dimmer galaxies and thus might be expected to be relatively small.

We chose to apply our luminosity cutoff firstly for the sake of completeness,
because we cannot see many of the dimmer spiral galaxies that must lie in 
our cosmic neighborhood (see Figure \ref{Vol-limit}). Additionally, we foresee our sample being used to
make comparisons with more distant samples, for instance, to study the evolution
of the SMBH. It seems likely that those distant samples will not
be able to observe these dim galaxies either. Providing a clear luminosity
limit may make such comparisons easier. Of course, ultimately we do aim
to study the extent to which these dimmer spirals do contribute to the BHMF, but there is an important caveat. It is by no means certain that
all such galaxies actually do contain black holes. They have been studied very little
and there have been claims that at least some such galaxies do not contain central black
holes, but only nuclear star clusters \citep{Ferrarese:2006}. For instance,
a large majority of the galaxies used to establish the $M$--$P$ relation 
in \citep{Berrier:2013} had a black hole with
mass greater than 6.5 million solar masses (the lowest mass SMBH in the sample use to define the $M$--$P$ relation was found in NGC 4395 with $\log(M/M_{\odot}) = 5.56_{-0.16}^{+0.12}$), so it clear that the relation
is much better constrained at the high-mass end than the low-mass end,
as with all other such relations.
Caution seems to be warranted in
exploring this part of the sample and we pass over it in this paper in
the face of such uncertainty.
 
\begin{figure} 
\includegraphics[trim = 4mm 5mm 0mm 0mm, clip, width=\columnwidth]{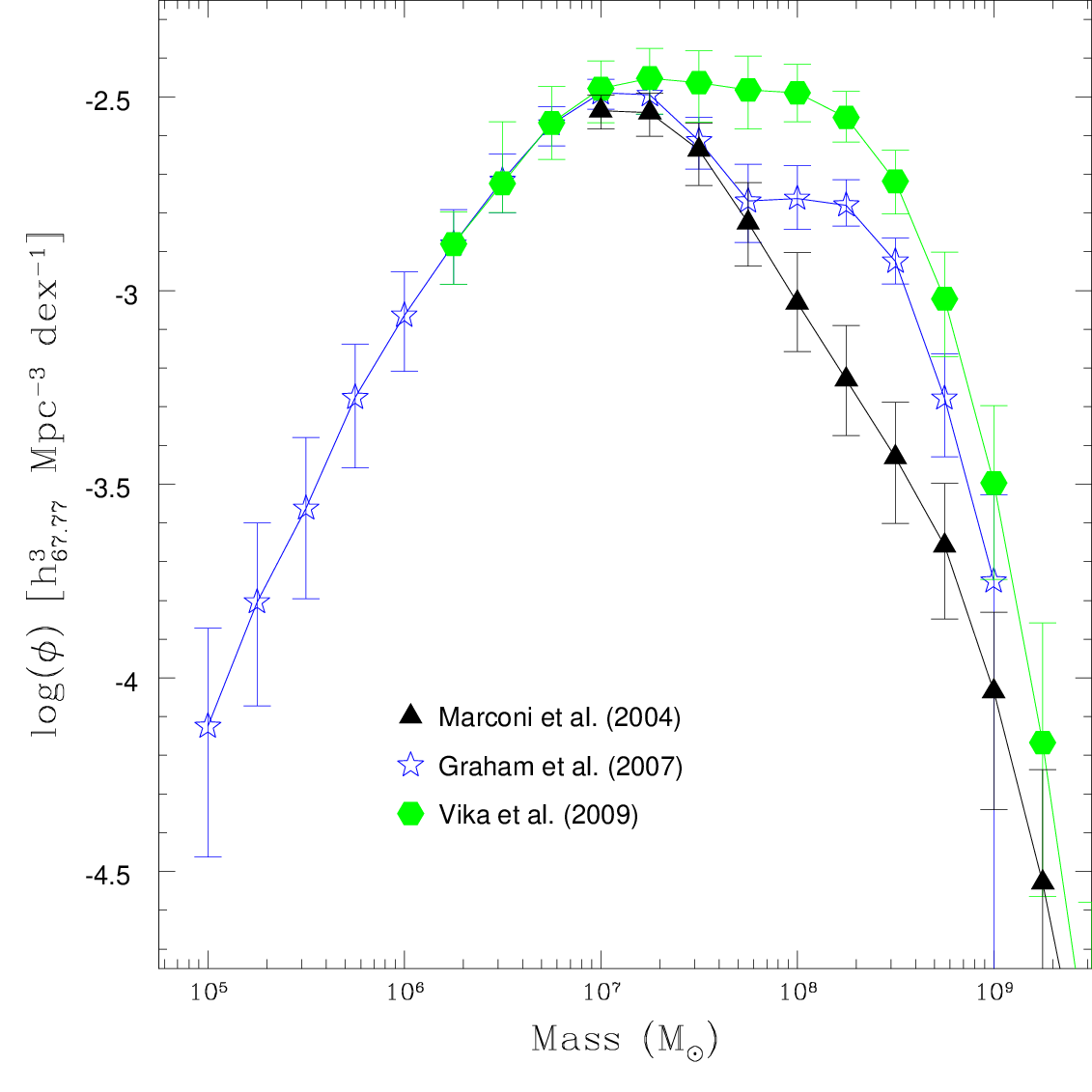}
\caption{Visualization of all-type BHMF mass functions generated by the addition of the MCMC PDF of our late-type BHMF with the early-type BHMFs of \citet{Marconi:2004}, \citet{Graham:2007}, and \citet{Vika:2009} represented by {\bf black} triangles, \textcolor{blue}{blue} stars, and \textcolor{green}{green} hexagons, respectively.\label{Plot_2}}
\end{figure}

Ultimately, a total BHMF for all types of galaxies is desired. In
Figure \ref{Plot_2}, 
we add the MCMC PDF of our late-type BHMF to the early-type BHMFs found
in \citet{Marconi:2004}, \citet{Vika:2009}, and \citet{Graham:2007}. It is of
particular interest to note that all three of these quite varied
sources (\citealt{Marconi:2004} uses $\sigma$, \citealt{Vika:2009} uses bulge luminosity, and \citealt{Graham:2007}
uses the S\'ersic index to derive their BHMFs) agree near the peak
of the BHMF, although there are considerable disagreements on the
high-mass end. This does suggest that if we could become more confident
of the true state of the late-type BHMF, then we would be in a position
to have a thorough understanding of the low-mass end of the all-type BHMF.


\section{Conclusions}\label{Conclusions}

Through the application of our established relationship between the mass of central SMBHs 
and the spiral arm pitch angle of their host galaxies \citep{Berrier:2013}, we have 
been able to establish a robust BHMF for SMBHs residing in spiral galaxies in the 
local universe. \citet{Berrier:2013} demonstrate that the $M$--$P$ relation has the 
lowest scatter of any method currently used to estimate the mass of SMBHs residing 
in spiral galaxies. Its strength resides in the relationship being statistically tight, 
relative ease of implementation, and its independence from cosmographic parameters. We have 
also ascertained the distribution of pitch angles in the local universe, finding that our 
Milky Way has a pitch angle slightly higher than the average nearby spiral galaxy. 
Intriguingly, the discovery that the most probable geometry of spiral arms is close 
to that of the Golden Spiral was a serendipitous result.

We have now implemented the first major use of the $M$--$P$ relation in this determination. 
We are encouraged that our estimate of the local mass density of SMBHs in late-type galaxies 
agrees within order of magnitude with other published values.\footnote{
For instance, consider the values for local SMBH mass density given by
\citet{Graham:2007} and \citet{Vika:2009}; $(9.1 \pm 4.6)$ $\times$ $10^4$ 
$h_{67.77}^3$ $M_{\odot}$ Mpc$^{-3}$ and $(8.7 \pm 1.8)$ $\times$ $10^4$ $h_{67.77}^3$ $M_{\odot}$ 
Mpc$^{-3}$, respectively. Additionally, we are in rough agreement with the cosmological SMBH mass 
densities given by \citet{Graham:2007} and \citet{Vika:2009}; $(6.8 \pm 3.9)$ $\times$ $10^{-7}$ 
$h_{67.77}$ and $(6.8 \pm 1.0)$ $\times$ $10^{-7}$ $h_{67.77}$, respectively. Furthermore, we 
are also in agreement with the universal baryonic fraction locked up in SMBHs residing in spiral 
galaxies estimated by \citet{Graham:2007} and \citet{Vika:2009}; $0.031_{-0.018}^{+0.017}$ 
$h_{67.77}^3$ $\permil$ and $0.031_{-0.005}^{+0.004}$ $h_{67.77}^3$ $\permil$, respectively.} 
Our generation of a pitch angle distribution function demonstrates that the most probable mass 
of a SMBH residing in a spiral galaxy is $\approx 1.16 \times 10^7 M_{\odot}$. This is 
approximately an order of magnitude less than the most probable mass of a SMBH residing 
in an early-type galaxy \citep{Marconi:2004}. Furthermore, our result is consistent with 
the current galactic evolutionary construct that galaxies evolve across the Hubble Sequence 
\citep{Hubble:1926} from late-type toward early-type galaxies.

The low-mass end of the BHMF presents a number of challenges.
Since high-mass black holes are found in more luminous galaxies, they are
naturally easier to study since data is easier to acquire. As long as
we are interested in local galaxies, this is not an insurmountable obstacle
in itself. We have, for the moment, not dealt with the dimmest class of
spiral galaxies, for the sake of completeness. Nevertheless, our sample is
still peaked at the region (from a million solar masses to 50 million
solar masses) that has been identified as the key region within which,
if we better understood the local SMBH function, we could learn more
about the accretion rates of quasars and AGN in the past. Specifically,
it would be possible to constrain the fractions of the Eddington accretion
rate at which low-mass or high-mass black holes had accreted in the past
\citep{Shankar:2009b}.

A natural assumption seen in early work on the continuity equation was that
all AGN accrete at the Eddington limit. Convenient though this would
be for modern astronomers, there is substantial evidence now that it is
untrue.
If we could assume that all black holes accrete at the same constant fraction
of their Eddington limit, then it would be easy to work out the 
evolution of
the BHMF. This is because each quasar luminosity observed would correspond to a given
mass of black hole. One could work out the mass and accretion rate of 
each black
hole and determine at what point in the local BHMF it would ultimately
appear. However, more realistically, suppose that there is 
a random distribution about a mean for each black hole, so that 
for instance, every black hole accretes at a set fraction of the Eddington limit
(the mean of the distribution) plus or minus some random amount
(determined by the width of the distribution). Then, it follows that some
large black holes will in fact accrete at a relatively small rate. When
they do, they can be mistaken for smaller black holes accreting
at the normal rate or better for a black hole of that size. The result is that if large
black holes often radiate at too small a rate, then we will tend to
overestimate the number of small black holes and their rate of accretion.
It is hard to tell the difference between large black holes underperforming
and small black holes over-performing. One way to check is to count the
number of small black holes that actually exist today. 

As discussed in the previous section, the quantity and results of studies on the
BHMF in spiral galaxies leaves much to be desired. We present ours as of particular interest because
it is a complete sample within set limits. As such it may prove easier for
future studies to compare their results to ours. Even if they have a broader
sample, it should be possible for them to compare with our sample within
our known limits. Of the other BHMF's available for comparison, it is encouraging
that we have good agreement, at the high-mass end,
with those of \citet{Marconi:2004} and \citet{Vika:2009}. This gives us confidence
that our numbers our generally reliable and we argue that we thus have the first
dependable estimate of the low-mass end of the spiral galaxy black home mass function.
Previous studies of the late-type mass function either have acknowledged problems with
spiral galaxies \citep{Graham:2007}, do not cover the low-mass end at all \citep{Marconi:2004},
or do so with a sample which is likely to suffer strongly from Malmquist bias and be
very incomplete for dimmer galaxies \citep{Vika:2009}. We hope that our sample will
thus be useful to those studying the evolution of the BHMF as a way
of constraining the final population of relatively low mass black holes in 
the universe. One important lesson already stands out. 
Previous estimates of
the low-mass end of the late-type mass function \citep{Graham:2007,Vika:2009} found evidence of 
far fewer low-mass black holes than we do. Studies
based on accretion models \citep[e.g.,][]{Shankar:2013} and semi-analytic
models \citep[e.g.,][]{Marulli:2008} have presented
results which suggest that at the
low-mass end of the local BHMF does not fall away at all from the maximum height 
reached at the high-mass end of the function. Thus there is no peak in
the BHMF according to these models, but rather a relatively flat 
(or even rising) distribution
from about $10^8$ solar masses downward in mass.
Our results are clearly far closer to these models than the earlier
observational results were. 
Therefore, there are grounds to be optimistic that ultimately
local measurements of the BHMF will be brought into line with efforts to
model its evolution from what is known of quasars in the past.

\acknowledgments

The authors gratefully acknowledge support for this work from 
NASA Grant NNX08AW03A. This  research has  made  use of  the 
NASA/IPAC Extragalactic  Database (NED) which  is operated by  
the Jet Propulsion  Laboratory,  California  Institute  of  Technology,  
under contract with the National Aeronautics and Space Administration. 
Cosmology calculations were performed using the {\it RED IDL} cosmology 
package, we thank its authors Leonidas and John Moustakas. We generate plots using {\it MATLAB} and {\it Supermongo}. Data from plots published in other papers was acquired using {\it Plot Digitizer}. We also thank 
the National Science Foundation for REU Site Grant No. 0851150, which 
contributed to a significant part of the data collection for this paper. 
Additionally, we thank an anonymous referee for comments that greatly 
improved the paper. This research has made use of NASA's Astrophysics 
Data System. We acknowledge the usage of the {\it HyperLeda} database 
(\url{http://leda.univ-lyon1.fr}).

\appendix

\setcounter{table}{0}
\renewcommand{\thetable}{A\arabic{table}}

\section{The Golden Spiral}\label{Golden Appendix}

The pitch angle for the Golden Spiral ($P_{\varphi}$) is determined by starting with the definition of a logarithmic spiral in polar coordinates
\begin{equation}
r = r_{0}e^{b\theta},
\label{Log_Spiral}
\end{equation}
where $r$ is the radius, $\theta$ is the central angle, $r_{0}$ is the initial radius when $\theta = 0^{\circ}$, and $b$ is a growth factor such that
\begin{equation}
b = \tan(P).
\label{growth_factor}
\end{equation}
The golden ratio ($\varphi$) of two quantities applies when the ratio of the sum of the quantities to the larger quantity ($A$) is equal to the ratio of the larger quantity to the smaller one ($B$), that is
\begin{equation}
\frac{A + B}{A} = \frac{A}{B} \equiv \varphi.
\label{Golden_Ratio}
\end{equation}
Solving algebraically, the golden ratio can be found via the only positive root of the quadratic equation with
\begin{equation}
\varphi = \frac{1 + \sqrt{5}}{2} = 1.6180339887...
\label{Golden_Ratio_Solution}
\end{equation}
The Golden Spiral is a unique logarithmic spiral, defined such that its radius grows every quarter turn in either direction ($\pm\pi/2$) by a factor of $\varphi$. A simple solution of the pitch angle of the Golden Spiral can be yielded first by application of Equation (\ref{Log_Spiral})
\begin{equation}
\varphi = e^{b(\pm\pi/2)},
\label{phi_equals}
\end{equation}
rearranging and taking the natural logarithm
\begin{equation}
\left | b \right | = \frac{\ln{\varphi}}{\pm\pi/2} = 0.3063489625...,
\label{b_rad}
\end{equation}
and finally application of Equation (\ref{growth_factor}) yields
\begin{equation}
\left | P_{\varphi} \right | = \arctan\left | b \right | \approx 17.03239113^{\circ}...
\label{Golden_Spiral_Pitch}
\end{equation}

Within the errors associated with the $M$--$P$ relation (Equation (\ref{M-P_Relation})), the associated mass
 of a SMBH residing in a spiral galaxy with pitch angle equal to that of the Golden Spiral and the most probable pitch angle from the PDF in Figure \ref{Pitch_Distribution} are equivalent with $\log(M/M_{\odot}) = 7.15 \pm 0.22$ and $\log(M/M_{\odot}) = 7.06 \pm 0.23$, respectively. Perhaps the most famous spiral galaxy, M51a or the ``Whirlpool" galaxy, also exhibits spiral arms close to the Golden Spiral with a pitch angle of $P = 16.26^{\circ} \pm 2.36^{\circ}$ \citep{Me:2012} and an implied SMBH mass of $\log(M/M_{\odot}) = 7.20 \pm 0.26$.

The Golden Spiral plays a significant role in both the history and lore of
 mathematics and art. It is closely approximated by the Fibonacci Spiral, which is
 not a true logarithmic spiral. Rather, it consists of a series of quarter-circle arcs whose
 radii are the consecutively increasing numbers of the Fibonacci Sequence. Both the Golden Ratio and Fibonacci Sequence are
 manifested in the geometry and growth rates of many structures in nature; both physical and
 biological. It is not surprising, therefore, that spiral galaxies should also have morphologies clustering about this aesthetically appealing case. Another situation similar in 
 superficial appearance occurs in
 cyclogenesis in planetary atmospheres (e.g., hurricanes). 
 This rate of radial growth is most familiar
 in the anatomical geometry of organisms. Well-known examples of roughly Golden 
 Spirals are found in
 the horns of some animals (e.g., rams) and belonging to the shells of mollusks such as the
 nautilus, snail, and a rare squid which retains its shell, {\it Spirula spirula}. Of course,
 spiral density waves are not required to have pitch angles close to the Golden Spiral.
 Their pitch angle depends on the ration of mass density in the disk (where the waves
 propagate) to the central mass. In the case of Saturn's rings, where this ration is far
 smaller than it is in disk galaxies, pitch angles are measured in tenths of degrees.
 The fact that spiral arms in galaxies happen to cluster about the aesthetically appealing
 example of the Golden Spiral may help explain the enduring fascination that images
 of spiral galaxies have had on the public for decades.
 
\section{The Milky Way}\label{MW Appendix}

Our own Milky Way has $m = 4$ and $|P| = 22.5^{\circ} \pm 2.5^{\circ}$, as measured from neutral hydrogen observations \citep{Levine:2006}. This implies a SMBH mass of $\log(M/M_{\odot}) = 6.82 \pm 0.30$ from the $M$--$P$ relation, compared to a direct measurement mass estimate from stellar orbits around {\it Sgr A*} \citep{Gillessen:2009} of $\log(M/M_{\odot}) = 6.63 \pm 0.04$. Although our Milky Way does not have a pitch angle close to the most probable pitch angle from our distribution, it is very similar to the mean pitch angle from Figure \ref{Pitch_Distribution} ($\mu = 21.44^{\circ}$), with an associated SMBH mass of $\log(M/M_{\odot}) = 6.88 \pm 0.25$. However, the mean of the black hole mass distribution from Figure \ref{Mass_Distribution} is even closer with $\log(M/M_{\odot}) = 6.72$. Our Milky Way is somewhat atypical in that it has four spiral arms, which is only the third most probable harmonic mode for a galaxy (see Figure \ref{Harmonic_Modes}).

\section{Sample}\label{Appendix Sample}

\begin{longtable*}{llrrrrcrccr} 
\tablecolumns{11}
\tablecaption{Volume-limited Sample\label{Sample_Table}}
\tablehead{
\colhead{Galaxy Name} & \colhead{Hubble Type} & \colhead{$B_{\rm T}$} & \colhead{$D_{\rm L}$ (Mpc)} & \colhead{$A_{\rm B}$} & \colhead{$\mathfrak{M}_B$} & \colhead{$m$} & \colhead{$P$ (deg)} & \colhead{Band} & \colhead{Image Source} & \colhead{$\log(M/M_{\odot})$}  \\
\colhead{(1)} & \colhead{(2)} & \colhead{(3)} & \colhead{(4)} & \colhead{(5)} & \colhead{(6)} & \colhead{(7)} & \colhead{(8)} & \colhead{(9)} & \colhead{(10)} & \colhead{(11)}
}
ESO 027-G001 & Sbc & 12.78 & 18.3 & 0.723 & $-19.26$ & 2 & $-15.67 \pm 5.30$ & 468.0 nm\tablenotemark{a} & 1 & $7.24 \pm 0.39$  \\
ESO 060-G019 & SBcd & 12.80 & 22.4 & 0.364 & $-19.31$ & 1 & $-6.20 \pm 1.63$ & $B$ & 4 & $7.83 \pm 0.20$  \\
ESO 097-G013 & Sb & 12.03 & 4.2 & 5.277 & $-21.37$ & 6 & $26.74 \pm 5.00$ & 790.4 nm\tablenotemark{b} & 10 & $6.55 \pm 0.42$  \\
ESO 121-G006 & Sc & 10.74 & 20.6 & 0.186 & $-21.01$ & \nodata & \nodata & \nodata & \nodata & \nodata  \\
ESO 138-G010 & Sd & 11.62 & 14.7 & 0.797 & $-20.01$ & 2 & $-43.68 \pm 10.10$ & 468.0 nm\tablenotemark{a} & 3 & $5.50 \pm 0.76$  \\
ESO 209-G009 & SBc & 12.44 & 15.0 & 0.935 & $-19.37$ & \nodata & \nodata & \nodata & \nodata & \nodata  \\
ESO 494-G026 & SABb & 12.63 & 11.1 & 1.528 & $-19.13$ & 2 & $29.49 \pm 3.91$ & 468.0 nm\tablenotemark{a} & 1 & $6.38 \pm 0.39$  \\
IC 1953 & Scd & 12.71 & 24.6 & 0.110 & $-19.36$ & 3 & $-14.21 \pm 1.98$ & $I$ & 8 & $7.33 \pm 0.24$  \\
IC 2051 & SBbc & 11.89 & 23.9 & 0.411 & $-20.41$ & 2 & $-10.38 \pm 2.43$ & $R$ & 4 & $7.57 \pm 0.24$  \\
IC 2163 & Sc & 12.00 & 24.7 & 0.314 & $-20.28$ & 4 & $21.10 \pm 4.54$ & 468.0 nm\tablenotemark{a} & 1 & $6.90 \pm 0.38$  \\
IC 2469 & SBab & 12.00 & 23.1 & 0.511 & $-20.33$ & \nodata & \nodata & \nodata & \nodata & \nodata  \\
IC 2554 & SBbc & 12.64 & 21.2 & 0.743 & $-19.73$ & 2 & $38.72 \pm 11.21$ & 565.0 nm\tablenotemark{c} & 1 & $5.81 \pm 0.79$  \\
IC 4402 & Sb & 12.06 & 19.0 & 0.403 & $-19.74$ & \nodata & \nodata & \nodata & \nodata & \nodata  \\
IC 4444 & SABb & 12.07 & 18.0 & 0.609 & $-19.82$ & 2 & $-31.50 \pm 2.06$ & 468.0 nm\tablenotemark{a} & 3 & $6.26 \pm 0.35$  \\
IC 4721 & SBc & 12.39 & 23.2 & 0.283 & $-19.72$ & 3 & $-6.55 \pm 0.23$ & 468.0 nm\tablenotemark{a} & 3 & $7.80 \pm 0.17$  \\
IC 4901 & SABc & 12.28 & 23.7 & 0.200 & $-19.79$ & 5 & $-15.57 \pm 1.93$ & H$\alpha$ & 6 & $7.24 \pm 0.24$  \\
IC 5240 & SBa & 12.69 & 25.4 & 0.054 & $-19.38$ & 2 & $-11.41 \pm 4.01$ & 468.0 nm\tablenotemark{a} & 3 & $7.50 \pm 0.31$  \\
IC 5325 & Sbc & 12.23 & 18.1 & 0.074 & $-19.13$ & 4 & $-19.98 \pm 6.77$ & 468.0 nm\tablenotemark{a} & 3 & $6.97 \pm 0.48$  \\
Milky Way & SBc & \nodata & 0.00833\tablenotemark{d} & \nodata & $-20.3$\tablenotemark{e} & 4 & $22.5 \pm 2.5$ & 21 cm & 11 & $6.82 \pm 0.30$  \\
NGC 134 & SABb & 11.26 & 18.9 & 0.065 & $-20.19$ & 3 & $28.54 \pm 6.61$ & 468.0 nm\tablenotemark{a} & 1 & $6.44 \pm 0.51$  \\
NGC 150 & SBb & 12.13 & 21.0 & 0.052 & $-19.54$ & 2 & $14.29 \pm 4.26$ & $B$ & 2 & $7.32 \pm 0.33$  \\
NGC 157 & SABb & 11.05 & 19.5 & 0.161 & $-20.56$ & 3 & $8.66 \pm 0.89$ & $B$ & 2 & $7.67 \pm 0.19$  \\
NGC 210 & SABb & 11.80 & 21.0 & 0.079 & $-19.89$ & 2 & $-15.81 \pm 3.25$ & 468.0 nm\tablenotemark{a} & 1 & $7.23 \pm 0.29$  \\
NGC 253 & SABc & 8.16 & 3.1 & 0.068 & $-19.39$ & 2 & $-20.78 \pm 7.71$ & $R$ & 4 & $6.92 \pm 0.54$  \\
NGC 255 & Sbc & 12.31 & 20.0 & 0.097 & $-19.29$ & 2 & $-13.14 \pm 6.57$ & 468.0 nm\tablenotemark{a} & 1 & $7.40 \pm 0.45$  \\
NGC 275 & SBc & 12.72 & 21.9 & 0.203 & $-19.19$ & \nodata & \nodata & \nodata & \nodata & \nodata  \\
NGC 289 & SBbc & 11.79 & 22.8 & 0.071 & $-20.07$ & 5 & $19.71 \pm 1.95$ & $B$ & 2 & $6.99 \pm 0.27$  \\
NGC 337 & SBcd & 12.12 & 22.1 & 0.407 & $-20.01$ & 3 & $-15.90 \pm 5.18$ & $B$ & 6 & $7.22 \pm 0.39$  \\
NGC 578 & Sc & 11.60 & 21.8 & 0.044 & $-20.14$ & 3 & $16.51 \pm 1.88$ & $B$ & 2 & $7.19 \pm 0.25$  \\
NGC 613 & Sbc & 10.99 & 25.1 & 0.070 & $-21.08$ & 3 & $21.57 \pm 1.77$ & $B$ & 2 & $6.87 \pm 0.27$  \\
NGC 685 & Sc & 11.75 & 15.2 & 0.083 & $-19.24$ & 3 & $15.71 \pm 4.67$ & 468.0 nm\tablenotemark{a} & 1 & $7.24 \pm 0.36$  \\
NGC 908 & SABc & 10.93 & 17.6 & 0.091 & $-20.39$ & 3 & $15.26 \pm 2.61$ & $B$ & 2 & $7.26 \pm 0.27$  \\
NGC 986 & Sab & 11.70 & 17.2 & 0.069 & $-19.54$ & 2 & $46.60 \pm 6.32$ & 468.0 nm\tablenotemark{a} & 1 & $5.32 \pm 0.60$  \\
NGC 988 & Sc & 11.42 & 17.2 & 0.098 & $-19.85$ & \nodata & \nodata & \nodata & \nodata & \nodata  \\
NGC 1068 & Sb & 9.46 & 13.5 & 0.122 & $-21.31$ & 2 & $20.61 \pm 4.45$ & 468.0 nm\tablenotemark{a} & 1 & $6.93 \pm 0.37$  \\
NGC 1084 & Sc & 11.61 & 21.2 & 0.096 & $-20.12$ & 2 & $15.74 \pm 2.15$ & $H$ & 12 & $7.23 \pm 0.25$  \\
NGC 1087 & SABc & 11.65 & 17.5 & 0.125 & $-19.69$ & 2 & $39.90 \pm 4.44$ & $R$ & 9 & $5.74 \pm 0.48$  \\
NGC 1097\tablenotemark{f} & SBb & 10.16 & 20.0 & 0.097 & $-21.45$ & 2 & $15.80 \pm 3.62$ & I & 2 & $7.23 \pm 0.31$  \\
NGC 1187 & Sc & 11.39 & 18.8 & 0.078 & $-20.06$ & 4 & $-21.96 \pm 3.61$ & $B$ & 2 & $6.85 \pm 0.34$  \\
NGC 1232 & SABc & 10.65 & 18.7 & 0.095 & $-20.80$ & 3 & $-25.71 \pm 5.43$ & $B$ & 2 & $6.62 \pm 0.44$  \\
NGC 1253 & SABc & 12.65 & 22.7 & 0.326 & $-19.46$ & 2 & $-19.71 \pm 7.66$ & 468.0 nm\tablenotemark{a} & 1 & $6.99 \pm 0.53$  \\
NGC 1255 & SABb & 11.62 & 21.5 & 0.050 & $-20.09$ & 3 & $13.09 \pm 2.57$ & 468.0 nm\tablenotemark{a} & 1 & $7.40 \pm 0.25$  \\
NGC 1300 & Sbc & 11.22 & 18.1 & 0.110 & $-20.17$ & 2 & $-12.71 \pm 1.99$ & $B$ & 2 & $7.42 \pm 0.23$  \\
NGC 1317 & SABa & 11.92 & 16.9 & 0.076 & $-19.30$ & 1 & $-9.12 \pm 1.41$ & 468.0 nm\tablenotemark{a} & 1 & $7.64 \pm 0.20$  \\
NGC 1325 & SBbc & 12.26 & 22.0 & 0.079 & $-19.53$ & 4 & $13.84 \pm 1.05$ & 468.0 nm\tablenotemark{a} & 1 & $7.35 \pm 0.21$  \\
NGC 1350 & Sab & 11.22 & 24.7 & 0.044 & $-20.43$ & 1 & $-20.57 \pm 5.38$ & 468.0 nm\tablenotemark{a} & 1 & $6.93 \pm 0.41$  \\
NGC 1353 & Sb & 12.41 & 24.4 & 0.118 & $-19.64$ & 4 & $13.68 \pm 2.31$ & $B$ & 2 & $7.36 \pm 0.25$  \\
NGC 1357 & Sab & 12.44 & 24.7 & 0.157 & $-19.68$ & 2 & $-16.16 \pm 3.48$ & 468.0 nm\tablenotemark{a} & 1 & $7.21 \pm 0.31$  \\
NGC 1365 & Sb & 10.32 & 17.9 & 0.074 & $-21.02$ & 2 & $-34.81 \pm 2.80$ & $B$ & 2 & $6.05 \pm 0.39$  \\
NGC 1367 & Sa & 11.56 & 23.3 & 0.089 & $-20.36$ & 2 & $32.90 \pm 5.92$ & 468.0 nm\tablenotemark{a} & 1 & $6.17 \pm 0.50$  \\
NGC 1385 & Sc & 11.52 & 15.0 & 0.073 & $-19.43$ & 3 & $35.83 \pm 5.43$ & 468.0 nm\tablenotemark{a} & 1 & $5.99 \pm 0.49$  \\
NGC 1398 & SBab & 10.53 & 21.0 & 0.049 & $-21.13$ & 4 & $19.61 \pm 3.07$ & $V$ & 2 & $6.99 \pm 0.30$  \\
NGC 1425 & Sb & 11.44 & 21.3 & 0.047 & $-20.24$ & 6 & $-27.70 \pm 4.78$ & 468.0 nm\tablenotemark{a}& 3 & $6.49 \pm 0.42$  \\
NGC 1433 & SBab & 10.76 & 10.0 & 0.033 & $-19.26$ & 6 & $-25.82 \pm 3.79$ & 468.0 nm\tablenotemark{a} & 3 & $6.61 \pm 0.37$  \\
NGC 1448 & Sc & 11.45 & 17.4 & 0.051 & $-19.80$ & 2 & $8.19 \pm 1.50$ & 468.0 nm\tablenotemark{a} & 3 & $7.70 \pm 0.20$  \\
NGC 1511 & Sa & 11.86 & 16.5 & 0.223 & $-19.45$ & \nodata & \nodata & \nodata & \nodata & \nodata  \\
NGC 1512 & Sa & 11.04 & 12.3 & 0.039 & $-19.45$ & 2 & $-7.00 \pm 1.45$ & 468.0 nm\tablenotemark{a} & 3 & $7.78 \pm 0.19$  \\
NGC 1515 & SABb & 11.92 & 16.9 & 0.051 & $-19.26$ & 1 & $-21.65 \pm 4.31$ & 468.0 nm\tablenotemark{a} & 1 & $6.87 \pm 0.37$  \\
NGC 1532 & SBb & 10.68 & 17.1 & 0.055 & $-20.53$ & \nodata & \nodata & \nodata & \nodata & \nodata  \\
NGC 1559 & SBc & 11.03 & 15.7 & 0.108 & $-20.05$ & 2 & $-26.61 \pm 9.69$ & $B$ & 2 & $6.56 \pm 0.67$  \\
NGC 1566 & SABb & 10.30 & 12.2 & 0.033 & $-20.17$ & 2 & $-17.81 \pm 3.67$ & $B$ & 2 & $7.11 \pm 0.32$  \\
NGC 1617 & SBa & 11.26 & 13.4 & 0.027 & $-19.40$ & 4 & $18.72 \pm 2.97$ & $B$ & 5 & $7.05 \pm 0.30$  \\
NGC 1640 & Sb & 12.38 & 19.1 & 0.125 & $-19.15$ & 4 & $22.12 \pm 8.13$ & 468.0 nm\tablenotemark{a} & 1 & $6.84 \pm 0.57$  \\
NGC 1672 & Sb & 10.33 & 14.5 & 0.085 & $-20.56$ & 2 & $18.22 \pm 14.07$ & 468.0 nm\tablenotemark{a} & 1 & $7.08 \pm 0.90$  \\
NGC 1703 & SBb & 12.06 & 17.4 & 0.121 & $-19.26$ & 2 & $19.30 \pm 5.15$ & $B$ & 4 & $7.01 \pm 0.40$  \\
NGC 1792 & Sbc & 10.82 & 13.2 & 0.082 & $-19.86$ & 3 & $-20.86 \pm 3.79$ & $B$ & 2 & $6.92 \pm 0.34$  \\
NGC 1808 & Sa & 10.76 & 11.6 & 0.110 & $-19.66$ & 2 & $23.68 \pm 7.77$ & 468.0 nm\tablenotemark{a} & 3 & $6.74 \pm 0.55$  \\
NGC 1832 & Sbc & 12.12 & 25.1 & 0.265 & $-20.15$ & 3 & $21.61 \pm 1.72$ & 468.0 nm\tablenotemark{a} & 7 & $6.87 \pm 0.27$ \\
NGC 1964 & SABb & 11.54 & 21.4 & 0.125 & $-20.24$ & 2 & $-12.86 \pm 3.49$ & $B$ & 2 & $7.41 \pm 0.29$  \\
NGC 2280 & Sc & 11.03 & 24.5 & 0.369 & $-21.29$ & 4 & $21.47 \pm 2.87$ & $B$ & 2 & $6.88 \pm 0.31$  \\
NGC 2397 & SBb & 12.85 & 22.7 & 0.743 & $-19.67$ & 6 & $-33.20 \pm 4.57$ & 468.0 nm\tablenotemark{a} & 1 & $6.15 \pm 0.44$  \\
NGC 2442 & Sbc & 11.34 & 17.1 & 0.734 & $-20.56$ & 2 & $14.95 \pm 4.20$ & $V$ & 2 & $7.28 \pm 0.33$  \\
NGC 2525 & Sc & 12.23 & 18.8 & 0.211 & $-19.36$ & 2 & $-23.09 \pm 11.12$ & H$\alpha$ & 8 & $6.78 \pm 0.74$  \\
NGC 2559 & SBc & 11.71 & 19.0 & 0.793 & $-20.48$ & 2 & $-25.26 \pm 14.93$ & $B$ & 5 & $6.64 \pm 0.97$  \\
NGC 2566 & Sb & 11.86 & 12.5 & 0.522 & $-19.15$ & 2 & $5.90 \pm 2.28$ & 468.0 nm\tablenotemark{a} & 1 & $7.84 \pm 0.22$  \\
NGC 2835 & Sc & 11.04 & 10.8 & 0.365 & $-19.50$ & 3 & $-23.97 \pm 2.22$ & $B$ & 2 & $6.72 \pm 0.30$  \\
NGC 2997 & SABc & 10.06 & 10.8 & 0.394 & $-20.50$ & 2 & $-38.16 \pm 10.53$ & 468.0 nm\tablenotemark{a} & 1 & $5.84 \pm 0.75$  \\
NGC 3059 & SBbc & 11.72 & 14.8 & 0.884 & $-20.02$ & 5 & $-22.77 \pm 5.20$ & $B$ & 5 & $6.80 \pm 0.41$  \\
NGC 3137 & SABc & 12.27 & 17.4 & 0.252 & $-19.18$ & 3 & $7.00 \pm 1.51$ & 468.0 nm\tablenotemark{a} & 1 & $7.78 \pm 0.20$  \\
NGC 3175 & Sab & 12.29 & 17.6 & 0.268 & $-19.21$ & 2 & $22.37 \pm 12.45$ & $R$ & 13 & $6.82 \pm 0.81$  \\
NGC 3511 & SABc & 11.53 & 14.3 & 0.247 & $-19.49$ & 2 & $28.21 \pm 2.27$ & 468.0 nm\tablenotemark{a} & 1 & $6.46 \pm 0.33$  \\
NGC 3521 & SABb & 9.73 & 12.1 & 0.210 & $-20.89$ & 6 & $21.86 \pm 8.41$ & $B$ & 14 & $6.85 \pm 0.58$  \\
NGC 3621 & SBcd & 10.10 & 6.8 & 0.291 & $-19.34$ & 2 & $-12.66 \pm 1.21$ & 468.0 nm\tablenotemark{a} & 1 & $7.43 \pm 0.21$  \\
\end{longtable*}
\clearpage
\addtocounter{table}{-1}
\begin{deluxetable*}{llrrrrcrccr}
\tablecolumns{11}
\tablecaption{(Continued)\label{Sample_Table}}
\tablehead{
\colhead{Galaxy Name} & \colhead{Hubble Type} & \colhead{$B_{\rm T}$} & \colhead{$D_{\rm L}$ (Mpc)} & \colhead{$A_{\rm B}$} & \colhead{$\mathfrak{M}_B$} & \colhead{$m$} & \colhead{$P$ (deg)} & \colhead{Band} & \colhead{Image Source} & \colhead{$\log(M/M_{\odot})$}  \\
\colhead{(1)} & \colhead{(2)} & \colhead{(3)} & \colhead{(4)} & \colhead{(5)} & \colhead{(6)} & \colhead{(7)} & \colhead{(8)} & \colhead{(9)} & \colhead{(10)} & \colhead{(11)}
}
\startdata
NGC 3673 & Sb & 12.62 & 24.8 & 0.203 & $-19.55$ & 5 & $19.34 \pm 4.58$ & $B$ & 4 & $7.01 \pm 0.37$  \\
NGC 3717 & Sb & 12.22 & 18.9 & 0.238 & $-19.40$ & \nodata & \nodata & \nodata & \nodata & \nodata  \\
NGC 3882 & SBbc & 12.80 & 20.2 & 1.404 & $-20.13$ & 2 & $19.30 \pm 2.69$ & 645.0 nm\tablenotemark{g} & 7 & $7.01 \pm 0.29$  \\
NGC 3887 & Sbc & 11.42 & 19.3 & 0.124 & $-20.13$ & 4 & $-29.16 \pm 4.82$ & $B$ & 2 & $6.40 \pm 0.43$  \\
NGC 3936 & SBbc & 12.83 & 22.6 & 0.293 & $-19.24$ & 2 & $17.21 \pm 3.40$ & 468.0 nm\tablenotemark{a} & 1 & $7.14 \pm 0.31$  \\
NGC 3981 & Sbc & 12.55 & 23.8 & 0.145 & $-19.48$ & 4 & $19.96 \pm 14.20$ & 468.0 nm\tablenotemark{a} & 1 & $6.97 \pm 0.91$  \\
NGC 4030 & Sbc & 11.67 & 24.5 & 0.096 & $-20.37$ & 3 & $23.48 \pm 5.76$ & $B$ & 2 & $6.75 \pm 0.44$  \\
NGC 4038 & SBm & 10.93 & 20.9 & 0.168 & $-20.84$ & 2 & $35.55 \pm 6.50$ & 468.0 nm\tablenotemark{a} & 1 & $6.01 \pm 0.54$ \\
NGC 4039 & SBm & 11.19 & 20.9 & 0.168 & $-20.58$ & 1 & $-14.38 \pm 5.37$ & 468.0 nm\tablenotemark{a} & 1 & $7.32 \pm 0.39$  \\
NGC 4094 & Sc & 12.51 & 20.8 & 0.205 & $-19.28$ & 3 & $14.96 \pm 4.82$ & 468.0 nm\tablenotemark{a} & 1 & $7.28 \pm 0.36$  \\
NGC 4219 & Sbc & 12.69 & 23.7 & 0.477 & $-19.66$ & 4 & $-26.50 \pm 6.96$ & 468.0 nm\tablenotemark{a} & 3 & $6.57 \pm 0.52$  \\
NGC 4487 & Sc & 12.21 & 20.0 & 0.077 & $-19.38$ & 2 & $28.27 \pm 9.02$ & $R$ & 9 & $6.46 \pm 0.63$  \\
NGC 4504 & SABc & 12.45 & 21.8 & 0.090 & $-19.33$ & 3 & $-28.26 \pm 4.23$ & 468.0 nm\tablenotemark{a} & 1 & $6.46 \pm 0.40$  \\
NGC 4594 & Sa & 9.08 & 10.4 & 0.186 & $-21.19$ & \nodata & \nodata & \nodata & \nodata & \nodata  \\
NGC 4666 & SABc & 11.80 & 18.2 & 0.090 & $-19.59$ & 4 & $25.34 \pm 4.49$ & 468.0 nm\tablenotemark{a} & 1 & $6.64 \pm 0.39$  \\
NGC 4699 & SABb & 10.56 & 24.7 & 0.125 & $-21.53$ & 5 & $17.72 \pm 3.97$ & $B$ & 5 & $7.11 \pm 0.33$  \\
NGC 4731 & SBc & 12.12 & 19.8 & 0.117 & $-19.47$ & 5 & $35.57 \pm 7.06$ & 468.0 nm\tablenotemark{a} & 1 & $6.00 \pm 0.57$  \\
NGC 4781 & Scd & 11.66 & 16.1 & 0.173 & $-19.55$ & 3 & $28.34 \pm 6.21$ & 468.0 nm\tablenotemark{a} & 1 & $6.45 \pm 0.49$  \\
NGC 4818 & SABa & 12.06 & 20.1 & 0.120 & $-19.57$ & 3 & $-25.14 \pm 5.28$ & 468.0 nm\tablenotemark{a} & 1 & $6.65 \pm 0.43$  \\
NGC 4835 & Sbc & 12.64 & 24.9 & 0.369 & $-19.71$ & 3 & $23.70 \pm 3.71$ & 468.0 nm\tablenotemark{a} & 1 & $6.74 \pm 0.35$  \\
NGC 4930 & Sb & 12.07 & 24.1 & 0.400 & $-21.08$ & 3 & $30.29 \pm 3.45$ & $B$ & 2 & $6.33 \pm 0.38$  \\
NGC 4941 & SABa & 12.05 & 18.2 & 0.132 & $-19.38$ & 4 & $20.42 \pm 3.37$ & $B$ & 5 & $6.94 \pm 0.32$  \\
NGC 4945 & SBc & 9.29 & 4.0 & 0.640 & $-19.34$ & \nodata & \nodata & \nodata & \nodata & \nodata  \\
NGC 4981 & Sbc & 12.33 & 24.7 & 0.153 & $-19.79$ & 3 & $20.47 \pm 11.66$ & $B$ & 1 & $6.94 \pm 0.76$  \\
NGC 5042 & SABc & 12.49 & 15.6 & 0.660 & $-19.14$ & 3 & $15.01 \pm 3.68$ & 468.0 nm\tablenotemark{a} & 3 & $7.28 \pm 0.31$  \\
NGC 5054 & Sbc & 11.85 & 19.9 & 0.299 & $-19.94$ & 3 & $-25.57 \pm 3.72$ & $B$ & 2 & $6.62 \pm 0.36$  \\
NGC 5121 & Sa & 12.47 & 25.2 & 0.259 & $-19.79$ & 2 & $-13.39 \pm 4.85$ & 468.0 nm\tablenotemark{a} & 3 & $7.38 \pm 0.36$  \\
NGC 5161 & Sc & 12.01 & 24.3 & 0.214 & $-20.13$ & 6 & $28.01 \pm 4.04$ & 468.0 nm\tablenotemark{a} & 3 & $6.47 \pm 0.39$  \\
NGC 5236 & Sc & 7.91 & 7.0 & 0.239 & $-21.54$ & 6 & $-16.04 \pm 1.74$ & $B$ & 2 & $7.22 \pm 0.24$  \\
NGC 5247 & SABb & 11.17 & 22.2 & 0.321 & $-20.88$ & 2 & $-31.94 \pm 5.75$ & $B$ & 2 & $6.23 \pm 0.49$  \\
NGC 5483 & Sc & 11.90 & 24.7 & 0.298 & $-20.36$ & 2 & $-22.98 \pm 4.52$ & $B$ & 2 & $6.79 \pm 0.38$  \\
NGC 5506 & Sab & 12.88 & 23.8 & 0.216 & $-19.22$ & \nodata & \nodata & \nodata & \nodata & \nodata  \\
NGC 5530 & SABb & 11.86 & 14.3 & 0.422 & $-19.34$ & 4 & $30.59 \pm 3.27$ & 468.0 nm\tablenotemark{a} & 3 & $6.31 \pm 0.38$  \\
NGC 5643 & Sc & 10.77 & 16.9 & 0.430 & $-20.80$ & 4 & $30.77 \pm 4.29$ & $B$ & 6 & $6.30 \pm 0.42$  \\
NGC 5713 & SABb & 12.09 & 23.8 & 0.142 & $-19.93$ & 2 & $-31.00 \pm 6.41$ & $R$ & 15 & $6.29 \pm 0.51$  \\
NGC 5792 & Sb & 12.52 & 24.4 & 0.210 & $-19.63$ & 2 & $16.77 \pm 7.95$ & 645.0 nm\tablenotemark{a} & 7 & $7.17 \pm 0.54$  \\
NGC 6118 & Sc & 12.30 & 23.4 & 0.571 & $-20.11$ & 2 & $13.63 \pm 6.09$ & 468.0 nm\tablenotemark{a} & 1 & $7.36 \pm 0.43$  \\
NGC 6215 & Sc & 11.99 & 20.5 & 0.599 & $-20.17$ & 4 & $-27.43 \pm 5.85$ & $B$ & 2 & $6.51 \pm 0.47$  \\
NGC 6221 & Sc & 10.77 & 12.3 & 0.598 & $-20.28$ & 6 & $-27.18 \pm 2.14$ & $B$ & 2 & $6.52 \pm 0.32$  \\
NGC 6300 & SBb & 11.01 & 15.1 & 0.353 & $-20.23$ & 4 & $-16.58 \pm 1.52$ & $B$ & 2 & $7.18 \pm 0.24$  \\
NGC 6744 & SABb & 9.13 & 9.5 & 0.155 & $-20.91$ & 5 & $21.28 \pm 3.80$ & 468.0 nm\tablenotemark{a} & 3 & $6.89 \pm 0.34$  \\
NGC 6814 & SABb & 12.30 & 22.8 & 0.664 & $-20.15$ & 4 & $26.05 \pm 6.48$ & $B$ & 16 & $6.59 \pm 0.49$  \\
NGC 7205 & Sbc & 11.64 & 19.4 & 0.082 & $-19.88$ & 4 & $-24.66 \pm 4.88$ & $B$ & 4 & $6.68 \pm 0.41$  \\
NGC 7213 & Sa & 11.71 & 22.0 & 0.055 & $-20.06$ & 4 & $7.05 \pm 0.28$ & 468.0 nm\tablenotemark{a} & 1 & $7.77 \pm 0.17$  \\
NGC 7218 & Sc & 12.50 & 24.8 & 0.119 & $-19.59$ & 4 & $18.53 \pm 3.57$ & 468.0 nm\tablenotemark{a} & 1 & $7.06 \pm 0.32$  \\
NGC 7314 & SABb & 11.68 & 18.5 & 0.078 & $-19.73$ & 5 & $22.23 \pm 2.60$ & $R$ & 4 & $6.83 \pm 0.30$  \\
NGC 7410 & SBa & 11.95 & 20.1 & 0.042 & $-19.61$ & 1 & $-5.63 \pm 2.42$ & $R$ & 4 & $7.86 \pm 0.23$  \\
NGC 7418 & Sc & 11.84 & 19.9 & 0.058 & $-19.71$ & 2 & $26.30 \pm 8.39$ & $R$ & 4 & $6.58 \pm 0.59$  \\
NGC 7531 & SABb & 11.89 & 22.8 & 0.038 & $-19.94$ & 2 & $18.31 \pm 9.06$ & $R$ & 4 & $7.07 \pm 0.61$  \\
NGC 7552 & Sab & 11.19 & 17.2 & 0.051 & $-20.03$ & 2 & $-15.08 \pm 4.21$ & $R$ & 4 & $7.28 \pm 0.33$  \\
NGC 7582 & SBab & 11.37 & 20.6 & 0.051 & $-20.25$ & 2 & $-14.66 \pm 7.44$ & $R$ & 4 & $7.30 \pm 0.51$  \\
NGC 7590 & Sbc & 12.11 & 25.3 & 0.062 & $-19.97$ & 5 & $-28.16 \pm 4.84$ & 468.0 nm\tablenotemark{a} & 3 & $6.46 \pm 0.42$  \\
NGC 7599 & SBc & 12.05 & 20.3 & 0.063 & $-19.55$ & 3 & $-27.89 \pm 7.72$ & $R$ & 4 & $6.48 \pm 0.56$  \\
NGC 7689 & SABc & 12.14 & 25.2 & 0.043 & $-19.91$ & 3 & $19.32 \pm 3.82$ & 468.0 nm\tablenotemark{a} & 3 & $7.01 \pm 0.33$ \\
NGC 7721 & Sc & 12.42 & 22.3 & 0.121 & $-19.44$ & 2 & $-21.55 \pm 2.59$ & $R$ & 4 & $6.87 \pm 0.30$  \\
NGC 7727 & SABa & 11.60 & 23.3 & 0.123 & $-20.36$ & 2 & $15.94 \pm 6.39$ & 468.0 nm\tablenotemark{a} & 1 & $7.22 \pm 0.45$  \\
PGC 48179 & SBm & 12.83 & 22.7 & 0.174 & $-19.12$ & 6 & $37.80 \pm 5.49$ & 468.0 nm\tablenotemark{a} & 1 & $5.87 \pm 0.51$
\enddata
\tablecomments{Columns: 
(1) Galaxy name.
(2) Hubble type, from {\it HyperLeda} \citep{Paturel:2003}.
(3) Total $B$-band apparent magnitude, from {\it HyperLeda} \citep{Paturel:2003}.
(4) Luminosity distance in Mpc, compiled from the mean redshift-independent distance from the NED. 
(5) Galactic extinction in the $B$-band from \cite{Schlafly:2011}, as compiled by the NED.
(6) $B$-band absolute magnitude, determined from the formula: $\mathfrak{M}_B = B_{\rm T} - 5\log(D_{\rm L}) + 5 - A_{\rm B} - K$, with $D_{\rm L}$ in units of $pc$ and $K$-corrections ($K$) set to zero for $z < 0.02$.
(7) Harmonic mode. 
(8) Pitch angle in degrees. 
(9) Filter waveband/wavelength used for pitch angle calculation. 
(10) Telescope/literature source of imaging used for pitch angle calculation. 
(11) SMBH mass in $\log(M/M_{\odot})$, converted from the pitch angle via Equation (\ref{M-P_Relation}).
Image Sources: 
(1) UK Schmidt (new optics); 
(2) \citet{Me:2012}; 
(3) UK 48 inch Schmidt; 
(4) ESO 1 m Schmidt; 
(5) CTIO 0.9 m; 
(6) CTIO 1.5 m; 
(7) Palomar 48 inch Schmidt; 
(8) OAN Martir 2.12 m; 
(9) La Palma JKT 1 m; 
(10) {\it HST}-WFPC2; 
(11) \citet{Levine:2006}; 
(12) 1.8 m Perkins; 
(13) MSSSO 1 m; 
(14) CTIO 4.0 m; 
(15) KP 2.1 m CFIM; 
(16) INT 2.5 m.} 
\tablenotetext{a}{IIIaJ emulsion.}
\tablenotetext{b}{F814W.}
\tablenotetext{c}{IIaD emulsion.}
\tablenotetext{d}{Distance estimate to the Galactic center from \citet{Gillessen:2009}.}
\tablenotetext{e}{$B$-band absolute magnitude from \citet{Kruit:1986}.}
\tablenotetext{f}{In addition to spiral arms in the disk of the galaxy, NGC 1097 displays rare $m = 2$ nuclear spiral arms in the bulge. These arms display an opposite chirality to the disk arms with $P = -30.60^{\circ} \pm 2.68^{\circ}$. If used, this would dictate a SMBH mass of $\log(M/M_{\odot}) = 6.31 \pm 0.36$.}
\tablenotetext{g}{103aE emulsion.}
\end{deluxetable*}
\clearpage
\bibliography{bibliography}

\end{document}